\documentclass[12pt,useAMS,useAASTEX,usenatbib]{aastex}
\usepackage{emulateapj5}

\makeatletter

\newenvironment{apjemufigure}{%
\def\@captype{figure}%
\noindent\begin{minipage}{0.999\linewidth}\begin{center}}
{\end{center}\end{minipage}}
\makeatother

\def\wm{{\sc wmap }}
\def\healpix{H{\sc ealpix }}
\def\glesp{G{\sc lesp }}
\def\meanchi{\overline{\chi^2}}
\def\brameanchi{\langle \meanchi \rangle}  
\def\wmap{\hbox{\sl WMAP~}}

\def\etal{et al.}

\def\alm{a_{\ell m}}
\def\Ylm{Y_{\ell m}}
\def\Cl{C_{\ell}}

\def\summ{\sum_{m=-\ell}^{\ell}}
\def\suml{\sum_{\ell=0}^{\infty}}
\def\lm{\ell m}
\def\a{{\bf a}}
\def\n{{\bf n}}
\def\Ph{{\bf \Psi}}
\def\F{{\bf \Phi}}
\def\Ks{{\bf \Xi}}
\def\ks{{\bf \xi}}
\def\G{{\bf G}}

\def\w{{\bf w}}
\def\I{{\bf I}}

\def\O{{\bf \Omega}}

\def\gam{{\bf \Gamma}}
\def\g{{\bf g}}

\def\planck{{\it Planck }}

\newcommand{\tac}{{Theoretical Astrophysics Center, Juliane Maries Vej
30, DK-2100,  Copenhagen, Denmark}}
\newcommand{\nbi}{{Niels Bohr Institute, Blegdamsvej 17,
DK-2100 Copenhagen, Denmark}}
\newcommand{\sao}{{Special Astrophysical Observatory, Nizhnij Arkhyz,
Karachaj-Cherkesia, 369167, Russia}}

\newcommand{\asc}{{Astro Space Center of Lebedev Physical Institute,
Profsoyuznaya 84/32, Moscow, Russia}}

\slugcomment{Submitted to {\it The Astrophysical Journal}}

\begin{document}

\title{Phase analysis of the 1-year \wmap data and its application for
 the CMB foreground separation}
 
\author{
Pavel D. Naselsky\altaffilmark{1,2},
Oleg V. Verkhodanov\altaffilmark{1,3},
Lung-Yih Chiang\altaffilmark{1},
Igor D. Novikov\altaffilmark{1,2,4}
}

\altaffiltext{1}{\tac}
\altaffiltext{2}{\nbi}
\altaffiltext{3}{\sao}
\altaffiltext{4}{\asc}

\email{naselsky@tac.dk, oleg@tac.dk}


\begin{abstract}
We present a new method based on phase analysis for the Galaxy and foreground
component separation from the cosmic microwave background (CMB)
signal. This method is based on a prevailing assumption that the phases of the
underlying CMB signal should have no or little correlation with those of the
foregrounds. This method takes into consideration all the phases of
each multipole mode ($\ell \le 50$, $-\ell \le m \le \ell$) from the
whole sky without galactic cut, masks or any dissection of the whole sky
into disjoint regions. We use cross correlation of the phases to
illustrate that significant correlations of the foregrounds manifest
themselves in the phases of the \wmap 5 frequency bands, which are
used for separation of the CMB from the signals. Our final
phase-cleaned CMB map has the angular power spectrum in agreement with
both the \wmap result and that from Tegmark, de Oliveira-Costa and
Hamilton (TOH), the phases of our derived CMB signal, however, are
slightly different from those of the \wmap Internal Linear Combination map and
the TOH map.   
\end{abstract}

\section{Introduction}
The release of the first year results from the Wilkinson Microwave
Anisotropy Probe (\wmap) opens a new epoch of the full-sky CMB
investigation, particularly for the analysis of different foreground
components \citep{wmap,wmapmap,wmapfg,wmapsystematics,wmappw}. The analyses 
from different groups, apart from the \wmap results, have raised
two main issues: firstly, why is the power of the quadrupole component
of the CMB signal considerably suppressed? \citep{toh,oliveira,efstathiou}
Secondly, is the CMB signal Gaussian?
\citep{wmapng,tacng,park,eriksen} The key point to these questions may
lie in the area of component separation, related with the removal of
the foregrounds from the all frequency maps.

The \wmap group produced an Internal Linear Combination (ILC) map
\citep{wmapfg} using a weighted combination of the 5 bands outside the
Galactic plane. \citet{toh} (hereafter TOH) perform an independent
foreground analysis from the \wmap data using weighting
coefficients dependent not only on the angular scales ($\ell$) but also on
the Galactic latitudes.
 
In this paper we propose an alternative method with the assumption
that the phases of the CMB signal derived from foreground cleaning should 
correlate as little as possible with those of the foregrounds.
According to the simplest inflation theory (see for review Komatsu et
al. 2003), pure CMB signals constitute a Gaussian random field, which
have random and uniformly distributed phases. Because of the
pronounced non-Gaussianity of the foregrounds, it therefore becomes
straightforward to regard the non-Gaussianity, if detected, from the \wmap
Internal Linear Combination (ILC) map, the TOH maps or any other derived
maps from the \wmap data as contaminations of the foregrounds at
various levels. As such, we should have significant cross-correlations
between the phases of the foregrounds and the CMB signal.
 
We perform the phase analysis of the signals for all the K--W bands of the
\wmap in order to remove the Galaxy and extragalactic foregrounds.
Below we will call our derived map the {\it Phase--Cleaned Map} (PCM), and
our method the PCM method. Different from other methods for
component separation, our method is based significantly on the
properties of phases, using each map from the \wmap K--W bands as
an image without any assumptions about the statistical nature of
the CMB signal. Moreover, we
use the whole sky for analyses and do not apply any Galaxy plane
cut-off and masks, nor do we dissect the whole sky into disjoint
regions. Our work is closely related with \citet{cc00,c3,nns,ccn} and
\citet{tacng}, from which some of the aspects of the CMB phase
analysis were developed. A new element is to implement the phases of
the maps from multifrequency K--W bands for reconstruction of the PCM.

The layout of this paper is as follows. In Section 2 we describe the
phases of the \wmap and illustrate the cross correlations between the
foreground maps provided in the \wmap website and between the 5
frequency maps. We introduce the ``non-blind'' PCM method
in Section 3 and in Section 4 we introduce the ``blind'' PCM method
and elaborate the 4 steps for foreground cleaning. We use a set of
simulated maps, which we call the \wm simulator as the numerical test
on the ``blind'' PCM method. We produce the PCM using the ``blind''
method in Section 7 and compare the PCM map and the power spectrum with
those of the \wmap ILC map and the TOH foreground-cleaned map and
Wiener-filtered map. We perform the Gaussianity test on these maps in
Section 8 and the Conclusion and Discussions are in Section 9.   

\section{The phases of the CMB and foregrounds}
The statistical characterization of the temperature fluctuations of
CMB plus foregrounds radiation and the instrumental noise on a sphere
can be expressed as a sum over spherical harmonics:
\begin{equation}
\Delta T(\theta,\varphi)=\suml \summ \alm \Ylm (\theta,\varphi),
\label{eq1}
\end{equation}
where $\alm$ are the coefficients of the expansion.
Homogeneous and isotropic CMB Gaussian random fields (GRFs), as a result
of the simplest inflation paradigm, possess Fourier modes whose real
and imaginary parts are Gaussian and mutually independent. The statistical
properties are then completely specified by its angular power spectrum
$\Cl^{cmb}$,  
\begin{equation}
\langle  a^{cmb}_{\ell^{ } m^{ }} (a^{cmb})^{*}_{\ell^{'} m^{'}}
\rangle = \Cl^{cmb} \; \delta_{\ell^{ } \ell^{'}} \delta_{m^{} m^{'}}.  
\label{eq2}
\end{equation}
In other words, from the Central Limit Theorem their phases
\begin{equation}
\Psi^{cmb}_{\ell m}=\tan^{-1}\frac{\Im (\alm^{cmb})}{\Re (\alm^{cmb})}
\label{eq3}
\end{equation}
are randomly and uniformly distributed at the range
$[0,2\pi]$. We hereafter denote at superscript  ``{\it cmb}''  as the
pure CMB signal. For the combined signal from the CMB and
foregrounds we can write down 
\begin{equation}
\alm=|\alm|\exp(i\Psi_{\lm}),
\label{eq4}
\end{equation}
where $|\alm|$ is the modulus and $\Psi_{\ell m}$ is the phase of each
$\alm$ harmonic. 

The statistical properties of the foreground signals are different from
those of GRFs. In the contemporary CMB experiments the measured
signals have contaminations from our Galaxy as well as from other
extragalactic sources at all frequency bands. For example,  for the
K--Ka bands  of the \wmap the foreground emissivity is mainly
determined by the combination of the synchrotron and
free-free emissions. For higher frequency bands (Q, V and W bands) the
contaminations include the free-free and dust components and some of
the residues from the synchrotron radiation.  Using different
assumption on the properties of the foregrounds such as the frequency
dependence for the amplitudes of the each kind of the foregrounds and
detector noise, the predicted spectral
index for the point--like sources, statistical independence of the
components exclude the Galaxy contaminations or the so-called
``blind'' Bayesian approach, several methods have been
proposed for the separation of the pure CMB signal from various foreground
contaminations
\citep{te96,bouchet,hobson,stolyarov,delabrouille,toh,patanchon,barreiro,gonzalez}.  
 
Our PCM method, complementary to the methods mentioned above, is based
on a prevailing assumption: the phases of the ``true'' CMB signal
should not correlate with those of the foregrounds. The advantage
of this assumption is that it allows us without any
special cut--offs of the Galactic plane to remove the foregrounds, in
particular the Galaxy contaminations from the \wmap K--W maps.
Below we use cross correlation of the phases between the \wmap bands
to demonstrate pronounced cross correlations of the foreground phases for
different bands, as shown in Fig~\ref{fig1}. All the phases of the foreground
components are derived from the \wmap foreground maps\footnote{\tt
http://lambda.gsfc.nasa.gov}.

\begin{apjemufigure}
\hbox{\hspace*{-0.5cm}
\centerline{\includegraphics[width=0.52\linewidth]{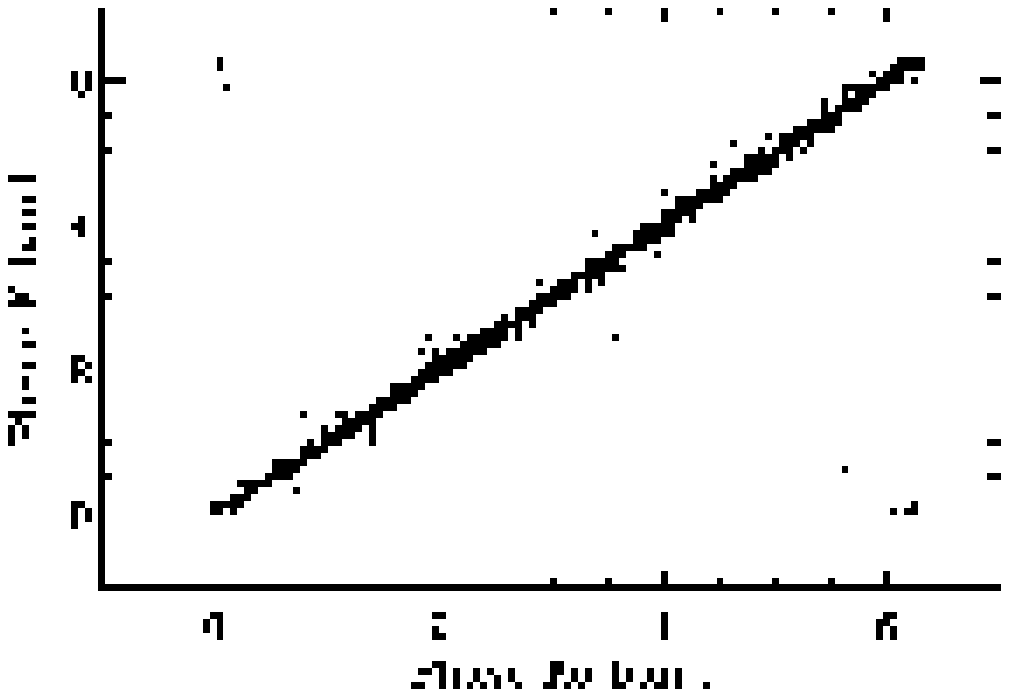}
\includegraphics[width=0.52\linewidth]{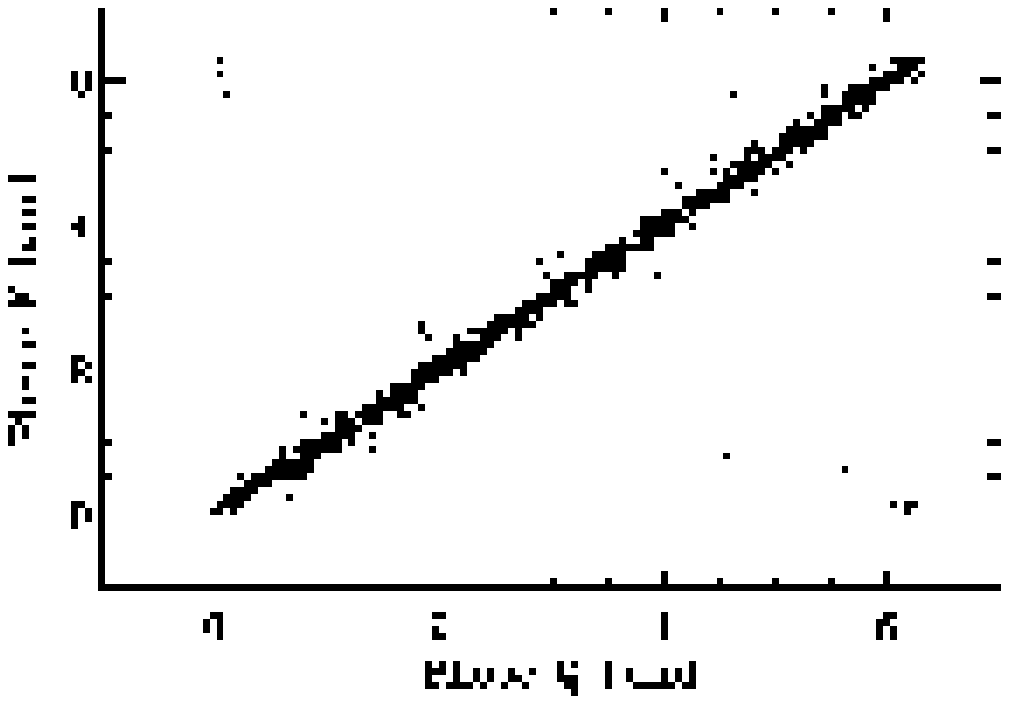}}}
\hbox{\hspace*{-0.5cm}
\centerline{\includegraphics[width=0.52\linewidth]{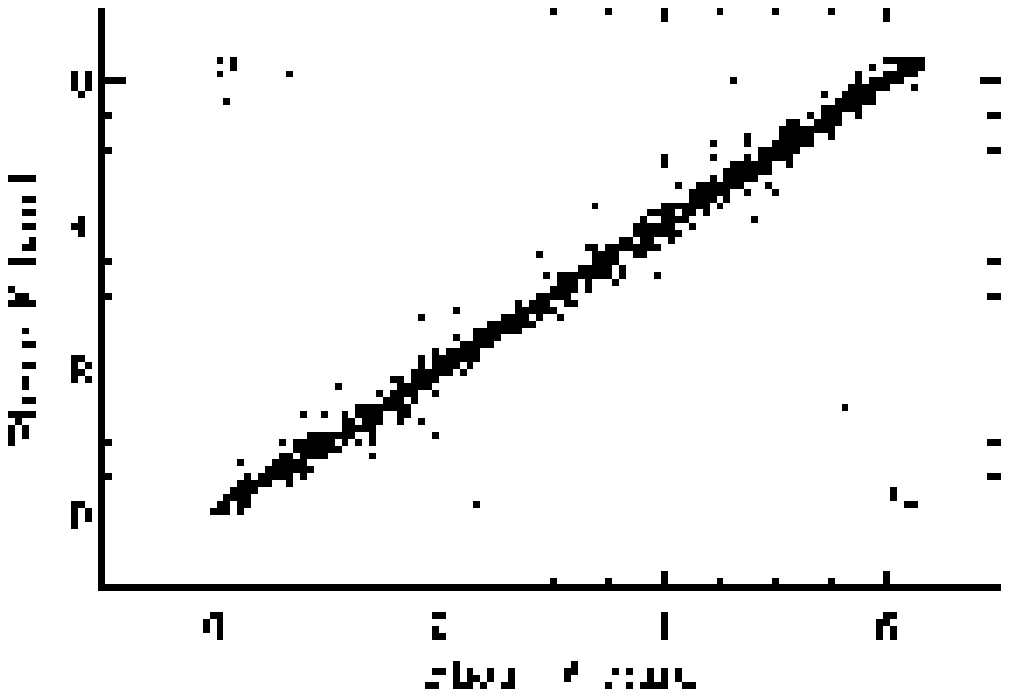}
\includegraphics[width=0.52\linewidth]{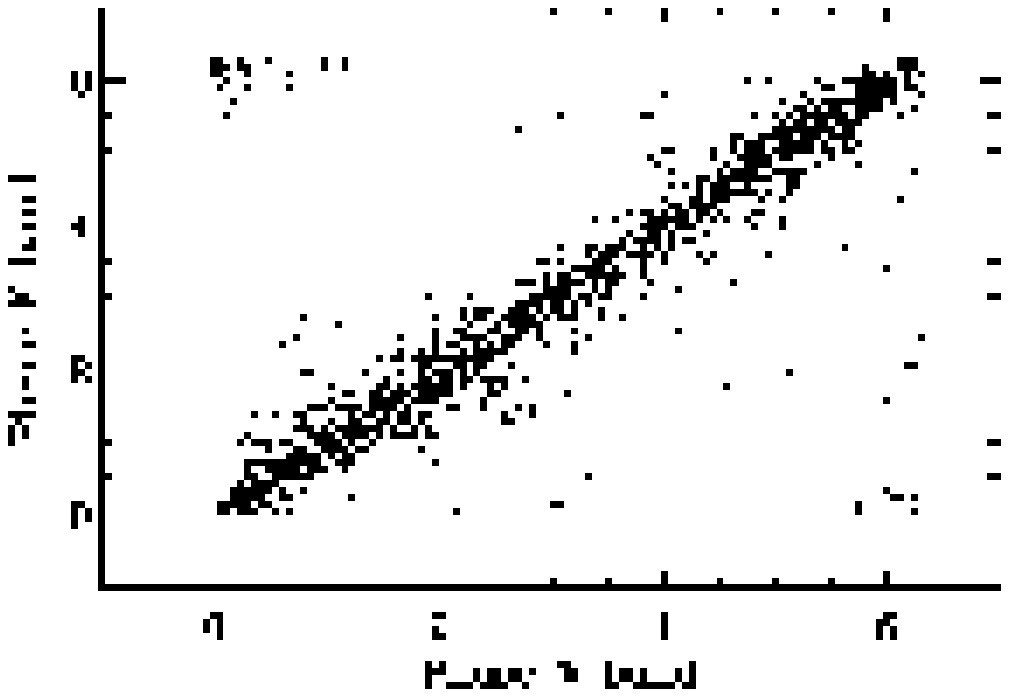}}}
\hbox{\hspace*{-0.5cm}
\centerline{\includegraphics[width=0.52\linewidth]{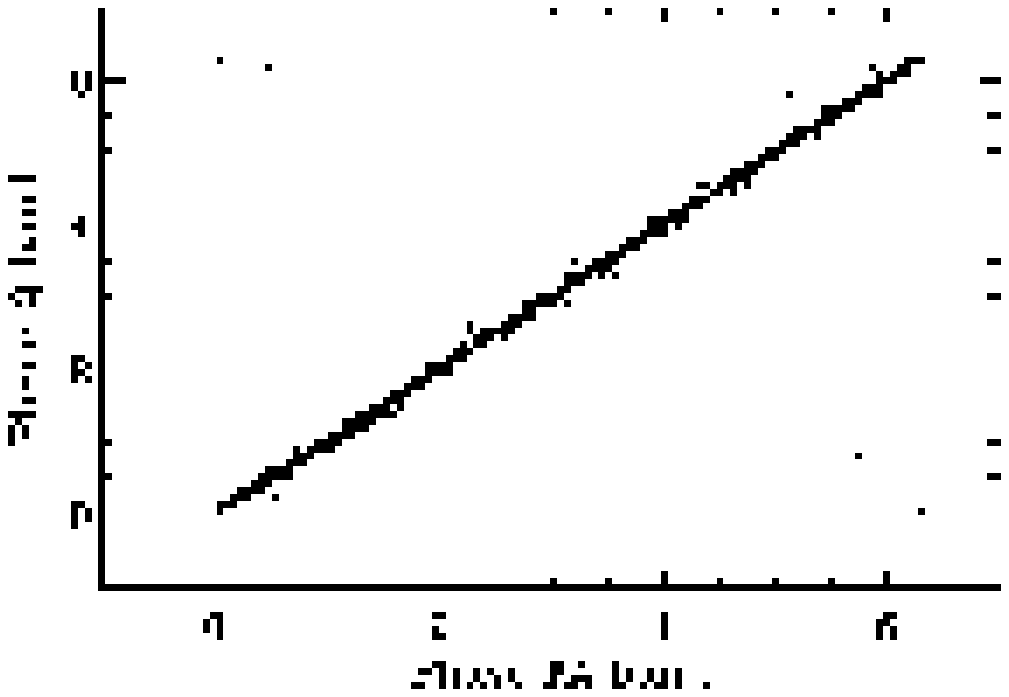}
\includegraphics[width=0.52\linewidth]{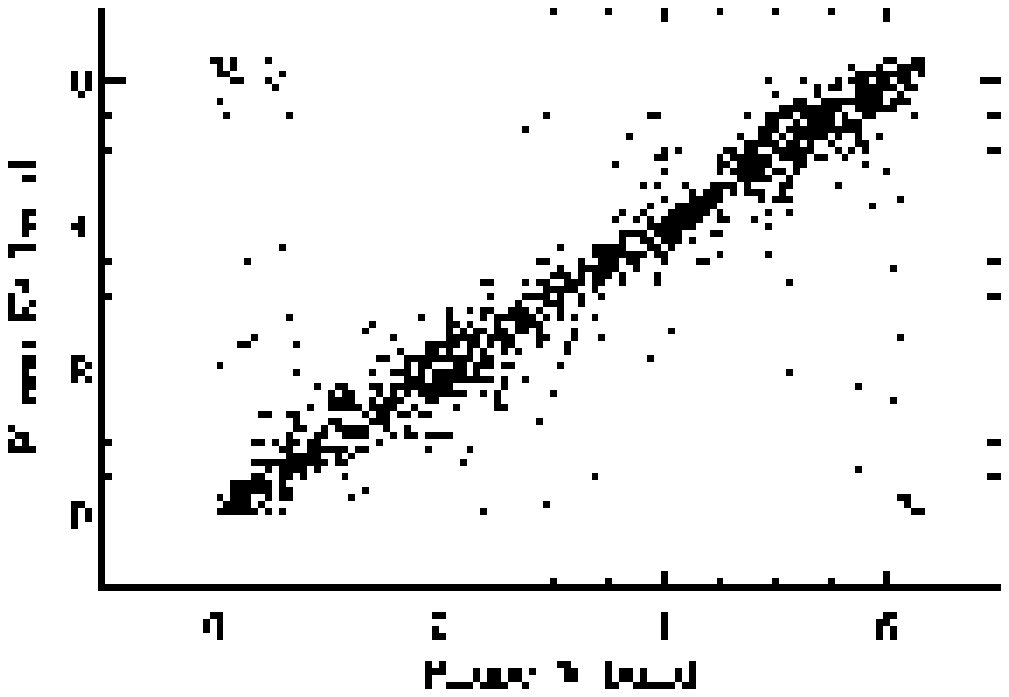}}}
\hbox{\hspace*{-0.5cm}
\centerline{\includegraphics[width=0.52\linewidth]{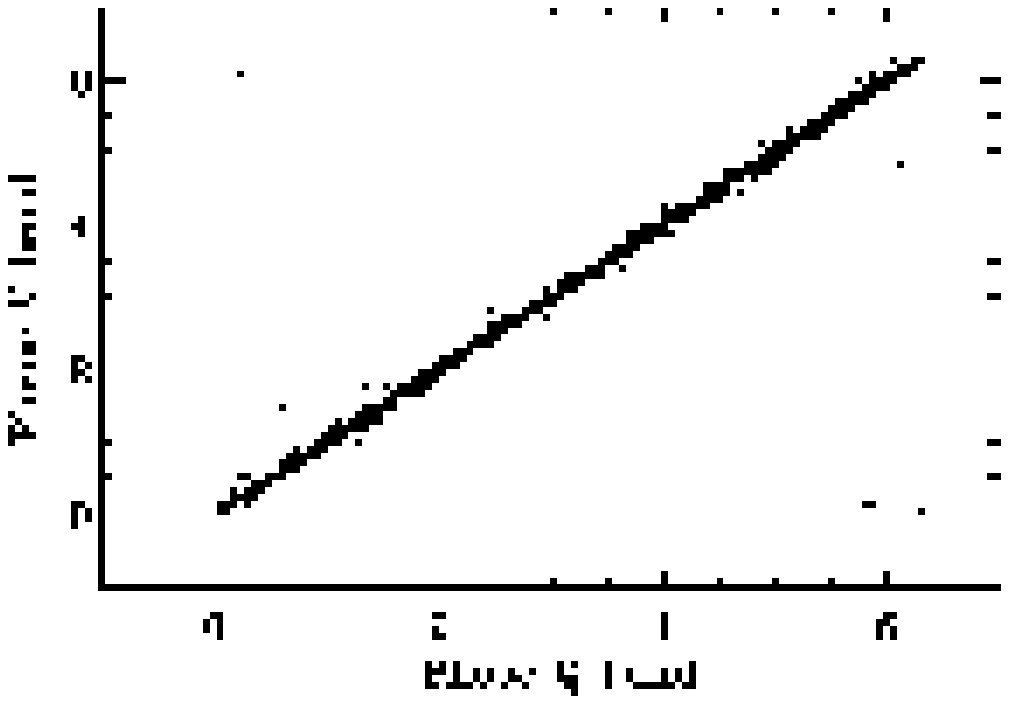}
\includegraphics[width=0.52\linewidth]{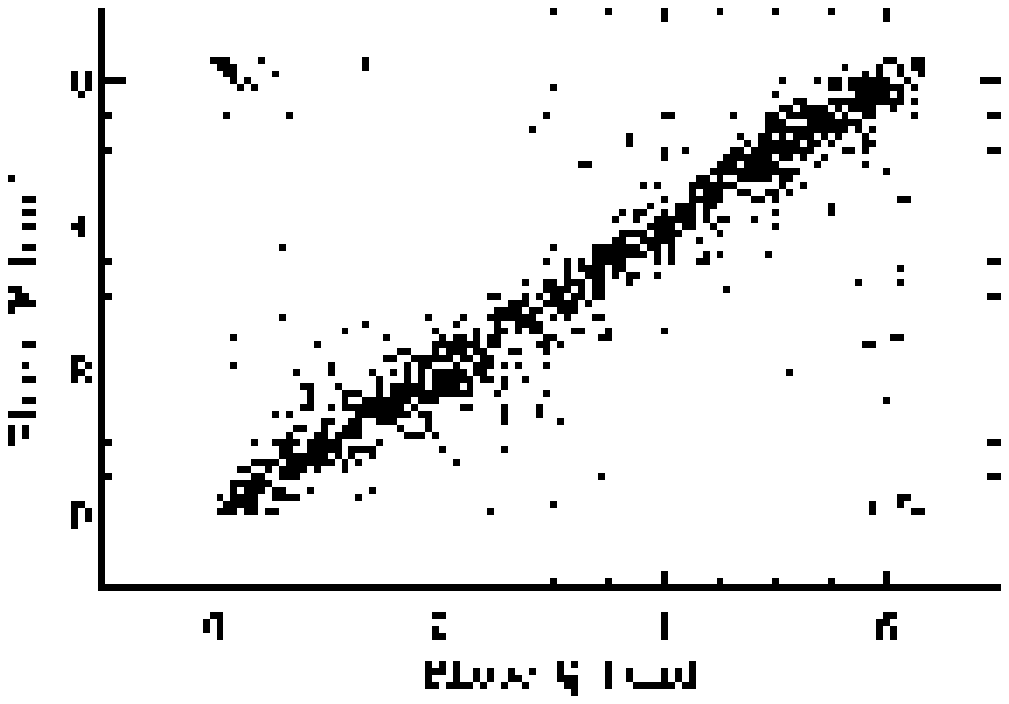}}}
\hbox{\hspace*{-0.5cm}
\centerline{\includegraphics[width=0.52\linewidth]{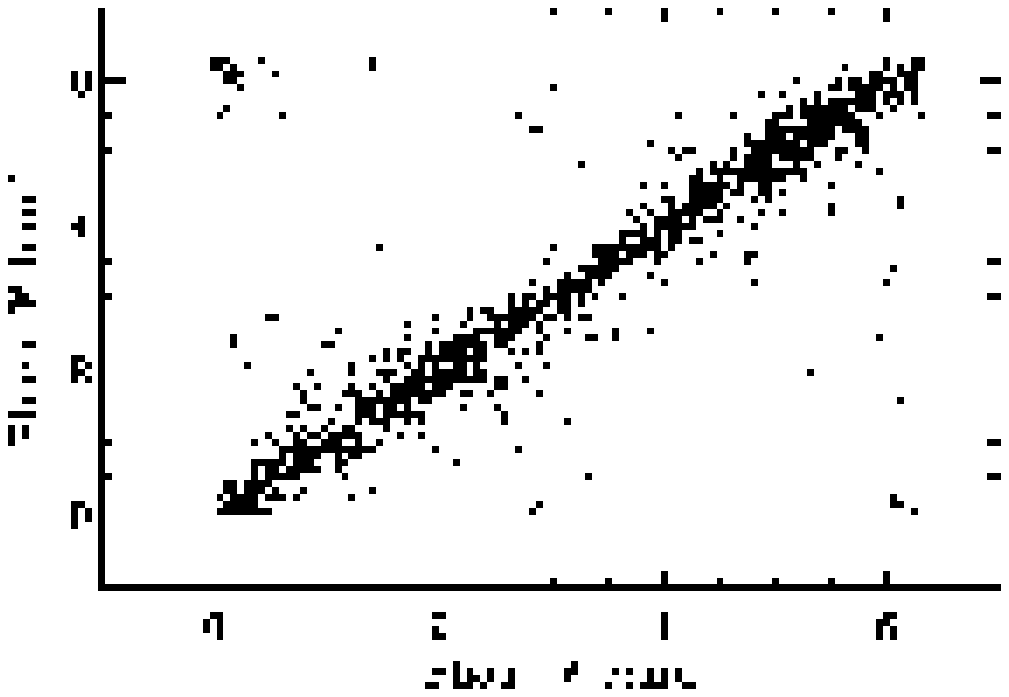}
\includegraphics[width=0.52\linewidth]{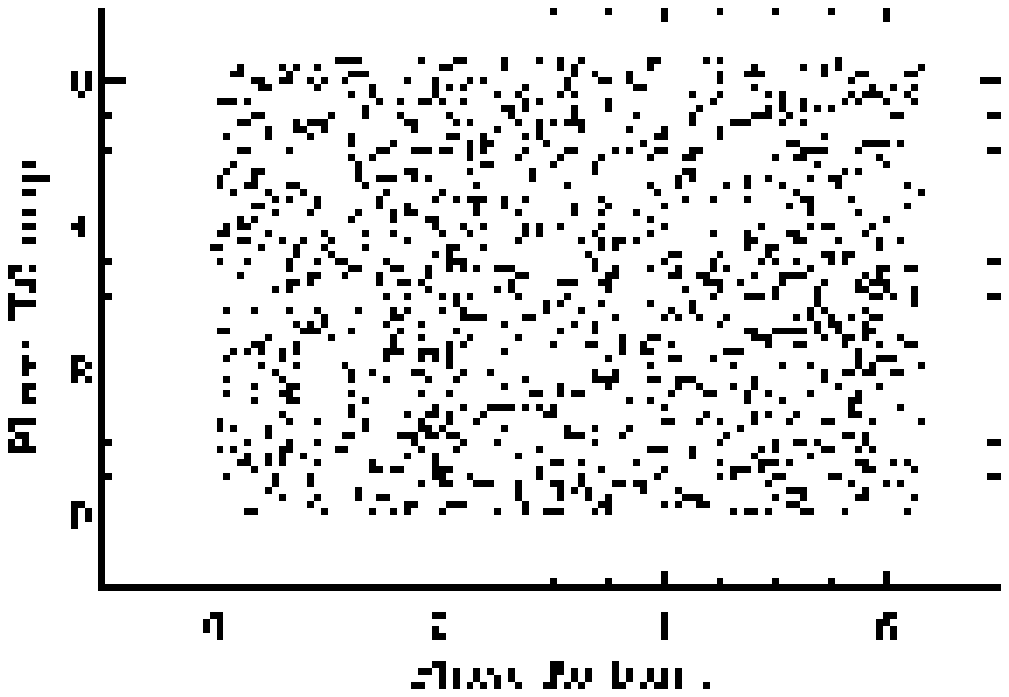}}}
\caption{The cross correlations of phases between all {\it foreground}
maps for the K--W bands. The phases are from all harmonic modes $\ell \le
50$. All the foreground maps are taken from the \wmap website. Note
also that on the bottom right panel we show the phases of the \wmap
ILC map and the foreground map at the Ka band, which displays no
correlations.}  
\label{fig1} 
\end{apjemufigure}

In Fig.\,2 we show the phase difference between the same $\ell,m$
modes but for different K--W foreground maps, e.g. $\Psi^{{\rm Ka},
f}_{\lm}-\Psi^{{\rm K},f}_{\lm}$ against $\Psi^{{\rm Ka},f}_{\lm}$.
One can see that the phase differences for Ka--Q band
and Q--V bands are minimal, unlike, for example, for Ka--W bands. For
comparison, in Fig.\,3 we plot the cross correlations of phases 
for the total phases between the bands. As one can see from Fig.\,1
and Fig.\,3, all the properties of the foregrounds phases correlations
manifest themselves in the cross correlation diagrams of the phases.

\begin{apjemufigure}
\hbox{\hspace*{-0.5cm}
\centerline{\includegraphics[width=0.52\linewidth]{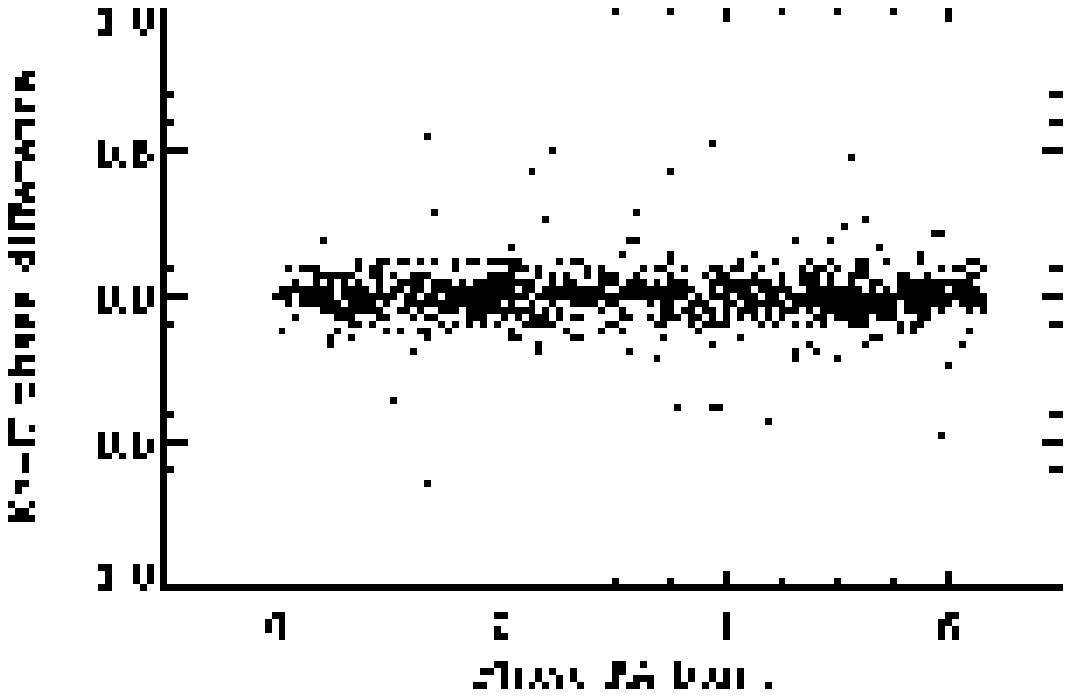}
\includegraphics[width=0.52\linewidth]{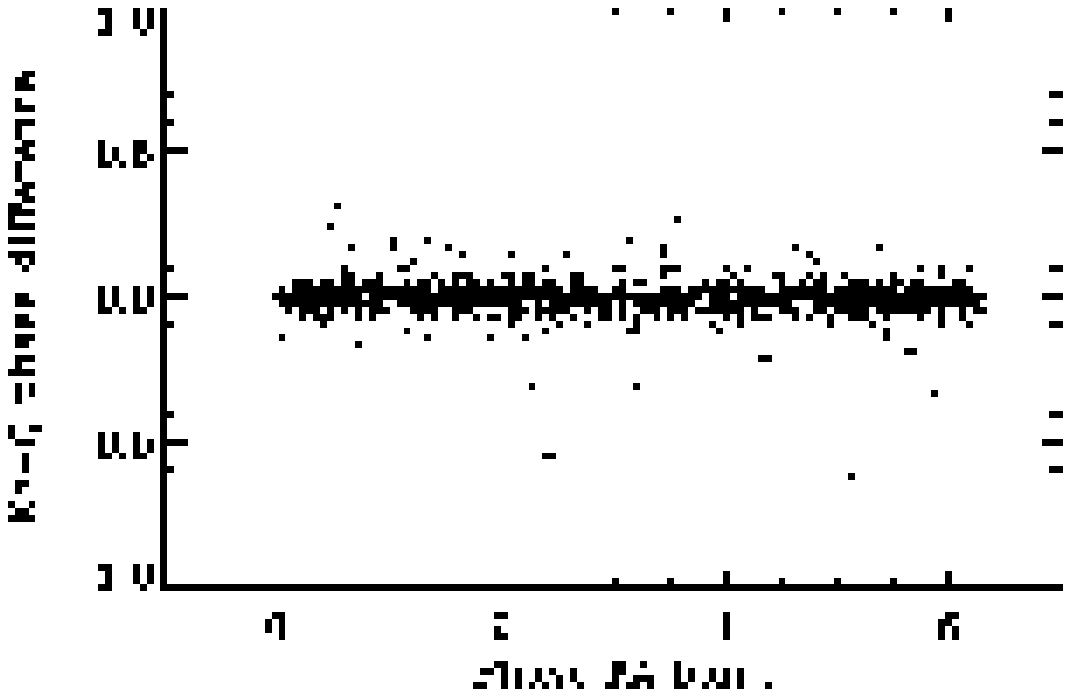}}}
\hbox{\hspace*{-0.5cm}
\centerline{\includegraphics[width=0.52\linewidth]{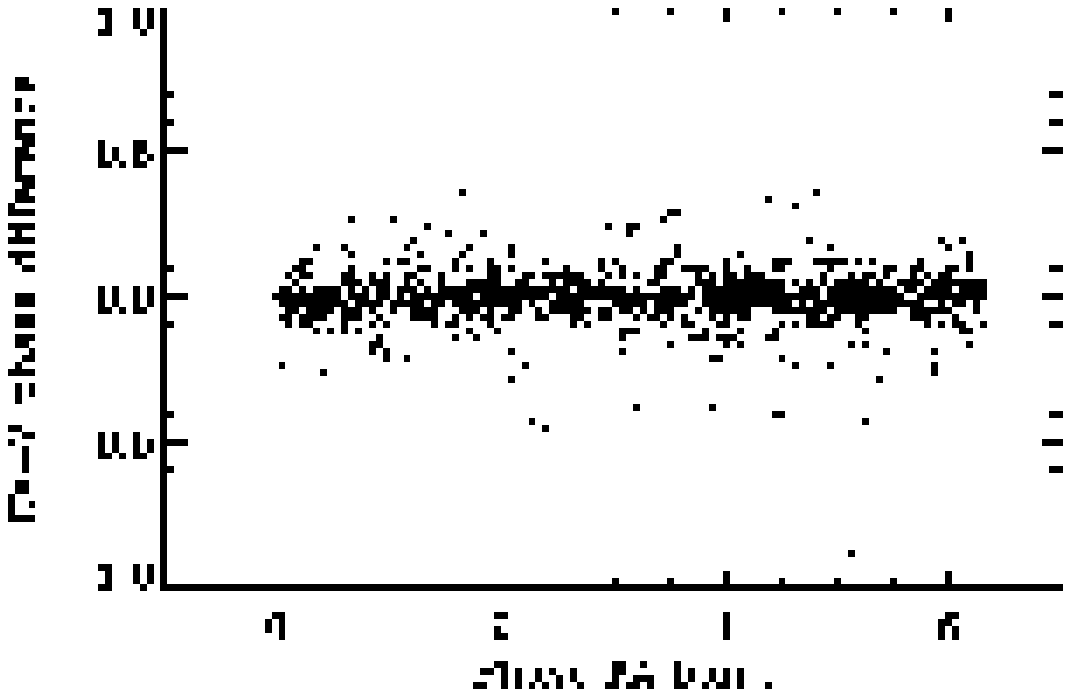}
\includegraphics[width=0.52\linewidth]{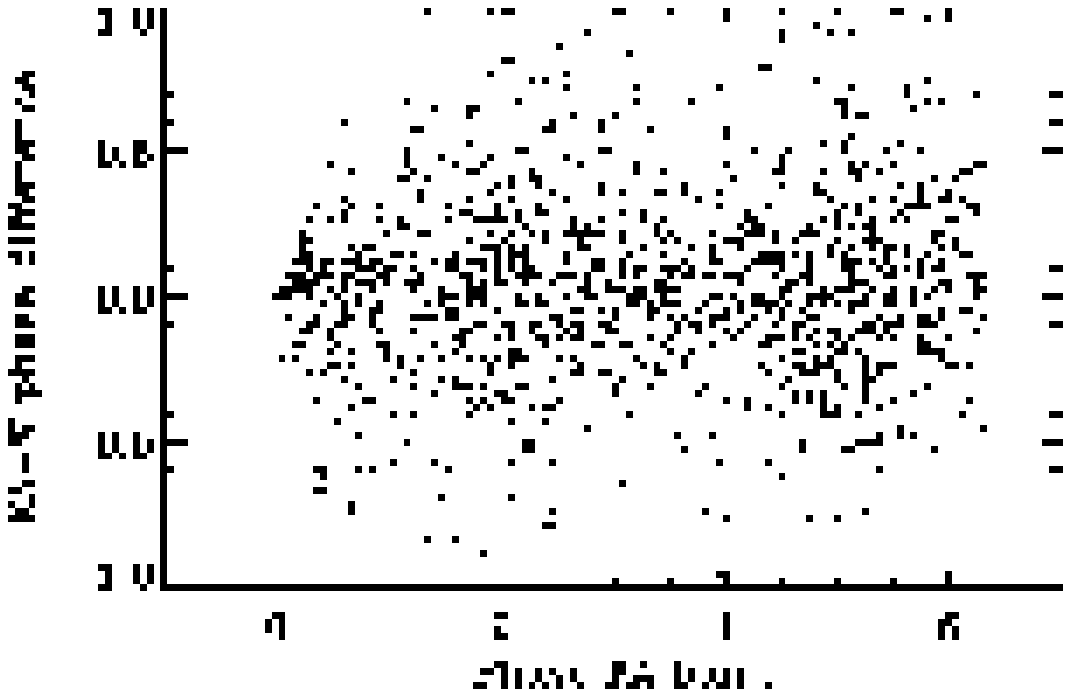}}}
\hbox{\hspace*{-0.5cm}
\centerline{\includegraphics[width=0.52\linewidth]{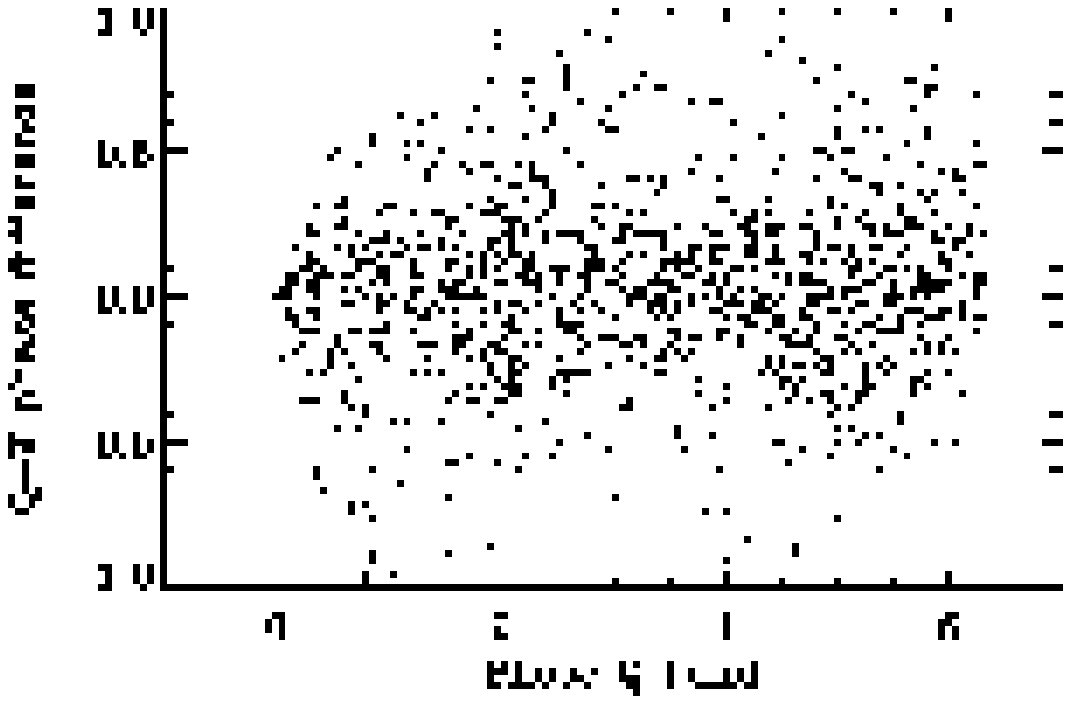}
\includegraphics[width=0.52\linewidth]{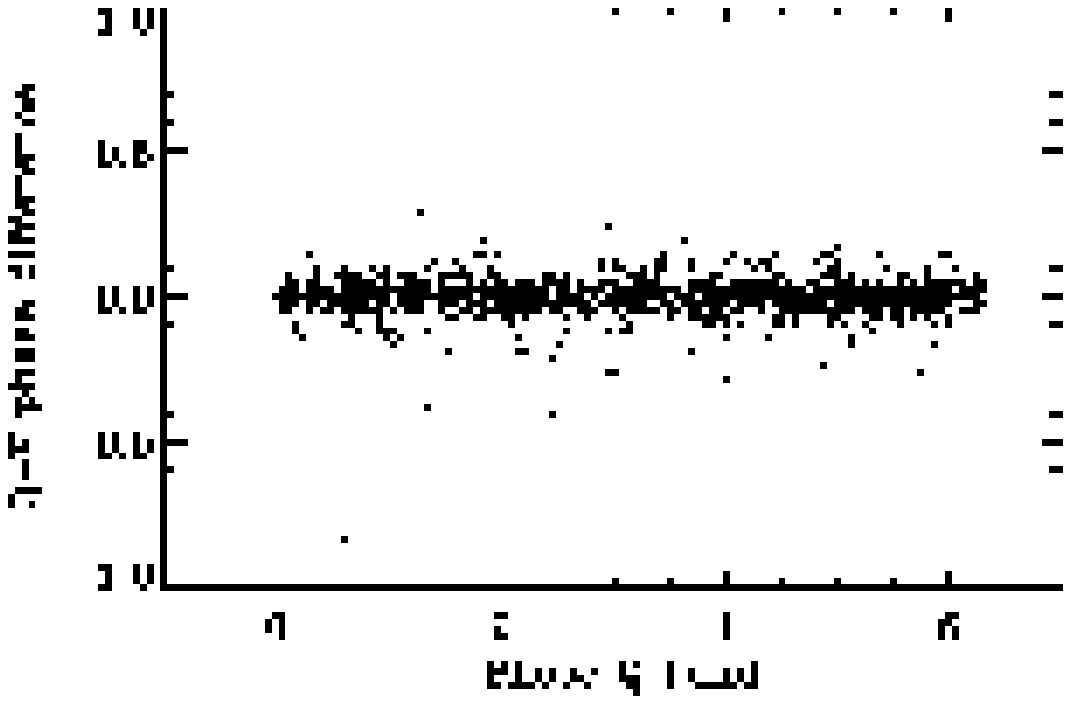}}}
\caption{The phase difference at the same $\ell,m$ modes for different
K--W foregrounds maps, e.g. $\Psi^{{\rm Ka},f}_{\lm}-\Psi^{{\rm K},f}_{\lm}$
against  $\Psi^{{\rm Ka},f}_{\lm}$ for the top left panel.}
\label{fig2} 
\end{apjemufigure}

In practice, we use the \healpix package \citep{healpix} to decompose
each of the
\wmap K--W maps for the coefficients of spherical harmonics $\a^{(j)}$, where
$\a \equiv \alm$ and hereafter the superscript index $j=1,2,3,4,5$
correspond to the K, Ka, Q, V, W bands, respectively, after which we use
the \glesp pixelization scheme \citep{glesp} for the rest of data processing in
order to control the accuracy of the $\a^{(j)}$ harmonics. We obtain the
phases $ \Ph^{(j)} \equiv \Psi^{(j)}_{\lm}$ from all  $\a^{(j)}$ (see
Eq.~(\ref{eq2})-(\ref{eq4})). Each set of the  $\a^{(j)}$ coefficients has
the combination of different kinds of the foregrounds  $\a^{(j),f}_k$
(convolved with the beam), the CMB signal $\a^{cmb}$ (also convolved
with the beam), and the instrumental noise $\n^{(j)}$:
\begin{equation}
\a^{(j)} = \sum_k \a^{(j),f}_k   + \a^{cmb} + \n^{(j)} \equiv
\G^{(j)}+ \a^{cmb},
\label{eq5}
\end{equation}
where $\sum_k \a^{(j),f}_k +\n^{(j)} \equiv \G^{(j)}$ is the common
signal excluding the CMB at the
$j$-th band. Unlike the linearity in Eq.(\ref{eq5}), the phases
of the $j$-th band are related to the phases of each kind of the foregrounds,
the CMB and the instrumental noise in a non-linear manner:  
\begin{equation}
\Ph^{(j)}=\tan^{-1}\frac{|\G^{(j)}|\sin{\Ks^{(j)}} +
 |\a^{cmb}|\sin{\ks}}{|\G^{(j)}|\cos{\Ks^{(j)}} + |\a^{cmb}|\cos{\ks}},
\label{eq6}
\end{equation}
where $\ks$ denotes the phases of the CMB signal and
\begin{eqnarray}
|\G^{(j)}|^2& = \sum_{k,q} \left[|\a^{(j),f}_k||\a^{(j),f}_q|
\cos(\Ph^{(j),f}_k- \Ph^{(j),f}_q) \right] + \nonumber \\
& |\n^{(j)}|^2 + 2 \sum_k  |\a^{(j),f}_k||\n^{(j)}|
 \cos(\Ph^{(j),f}_k-\F^{(j)}), 
\label{eq7}
\end{eqnarray}
where $\Ph^{(j),f}_q$ denotes the phases of the $q$-th foreground
component at the $j$-th band and $\F^{(j)}$ the phases of
the noise. The phases of the common signal excluding the CMB for
 the $j$-th band can be written as 
\begin{equation}
\Ks^{(j)}=\tan^{-1}\frac{\sum_k |\a^{(j),f}_k| \sin{\Ph^{(j),f}_k}
 +|\n^{(j)}|\sin{\F}^{(j)}}{ \sum_k |\a^{(j),f}_k|\cos{\Ph^{(j),f}_k}
 +|\n^{(j)}|\cos{\F^{(j)}}}. \label{eq8}
\end{equation}

\begin{apjemufigure}
\hbox{\hspace*{-0.5cm}
\centerline{\includegraphics[width=0.52\linewidth]{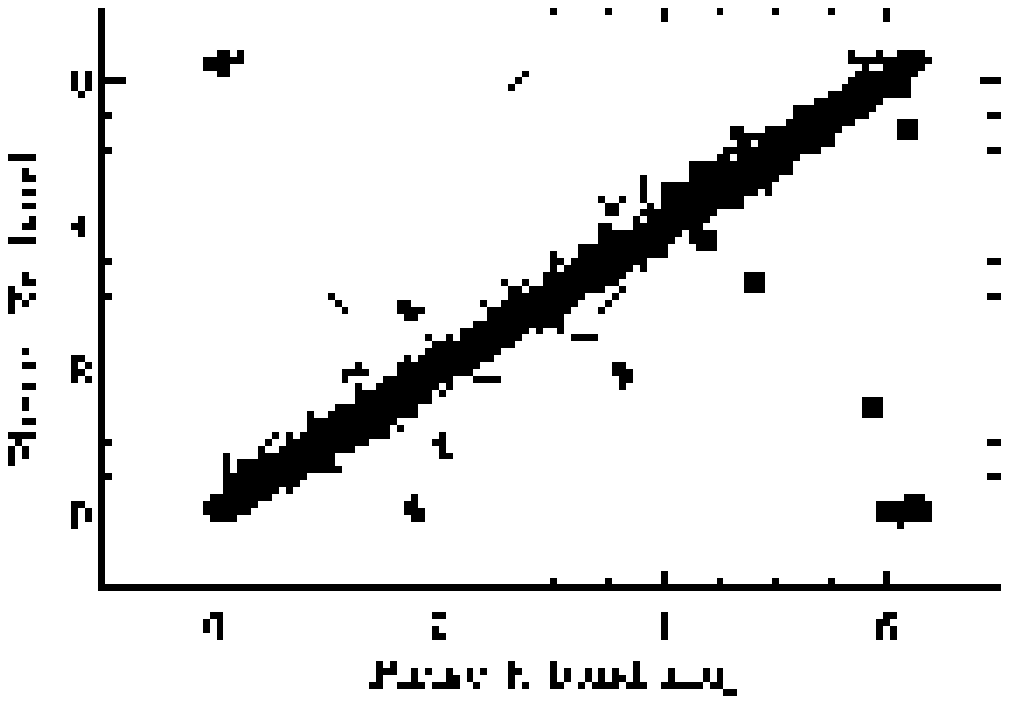}
\includegraphics[width=0.52\linewidth]{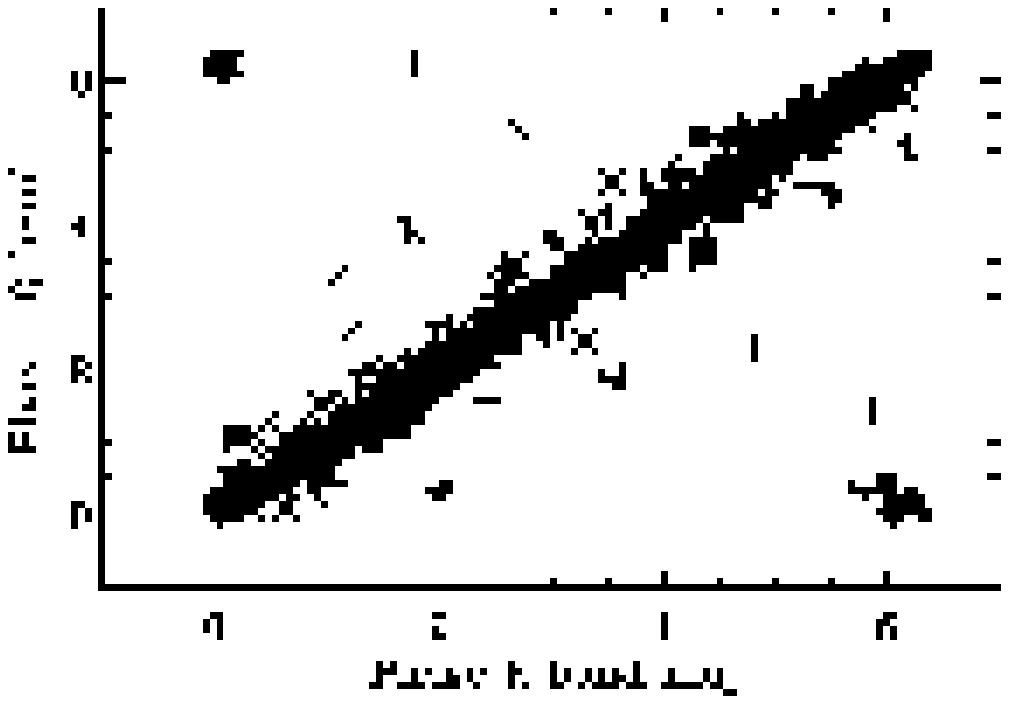}}}
\hbox{\hspace*{-0.5cm}
\centerline{\includegraphics[width=0.52\linewidth]{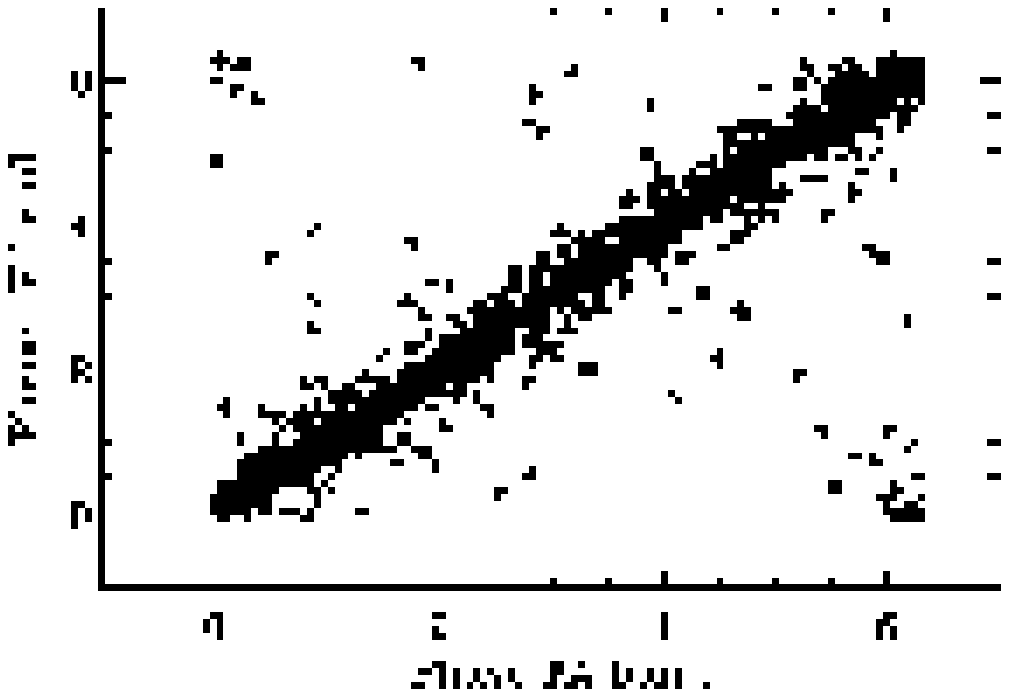}
\includegraphics[width=0.52\linewidth]{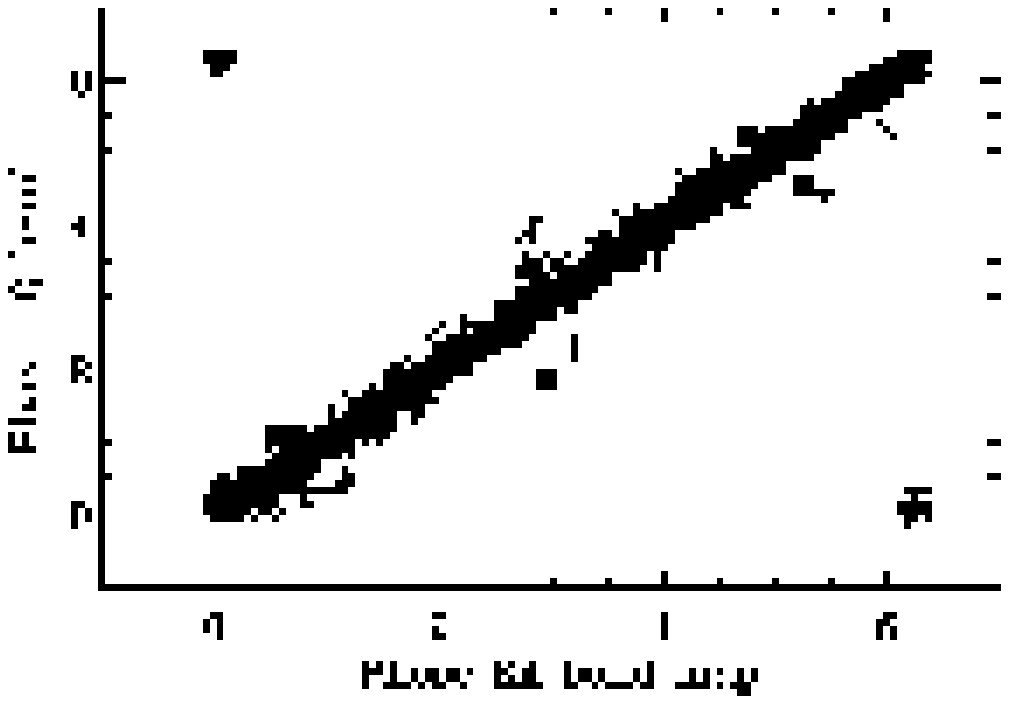}}}
\hbox{\hspace*{-0.5cm}
\centerline{\includegraphics[width=0.52\linewidth]{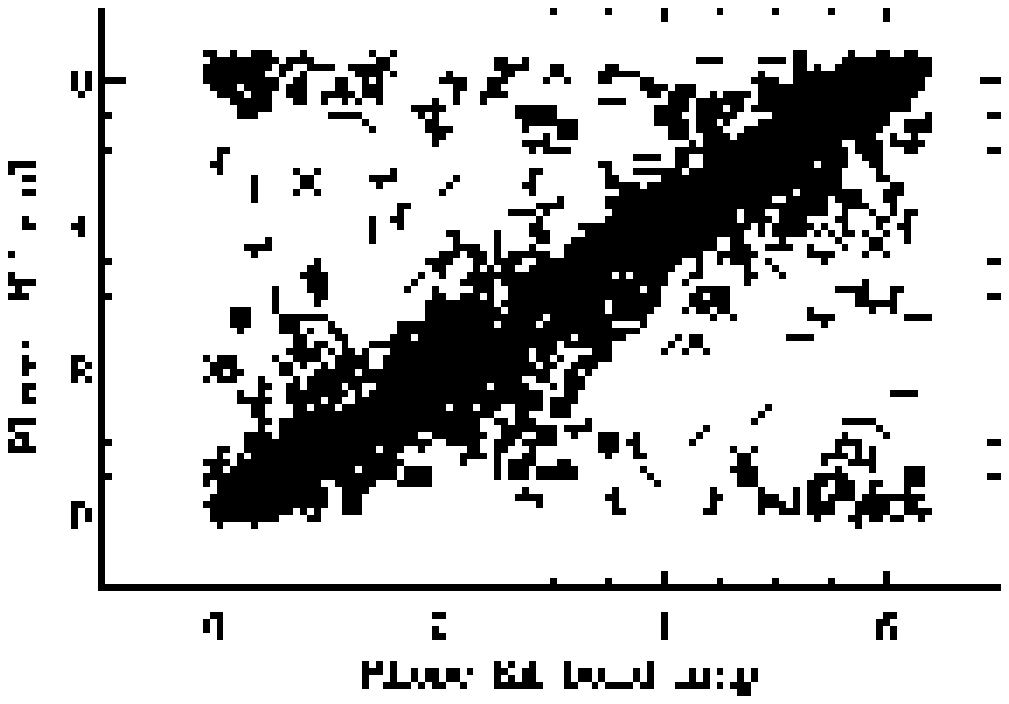}
\includegraphics[width=0.52\linewidth]{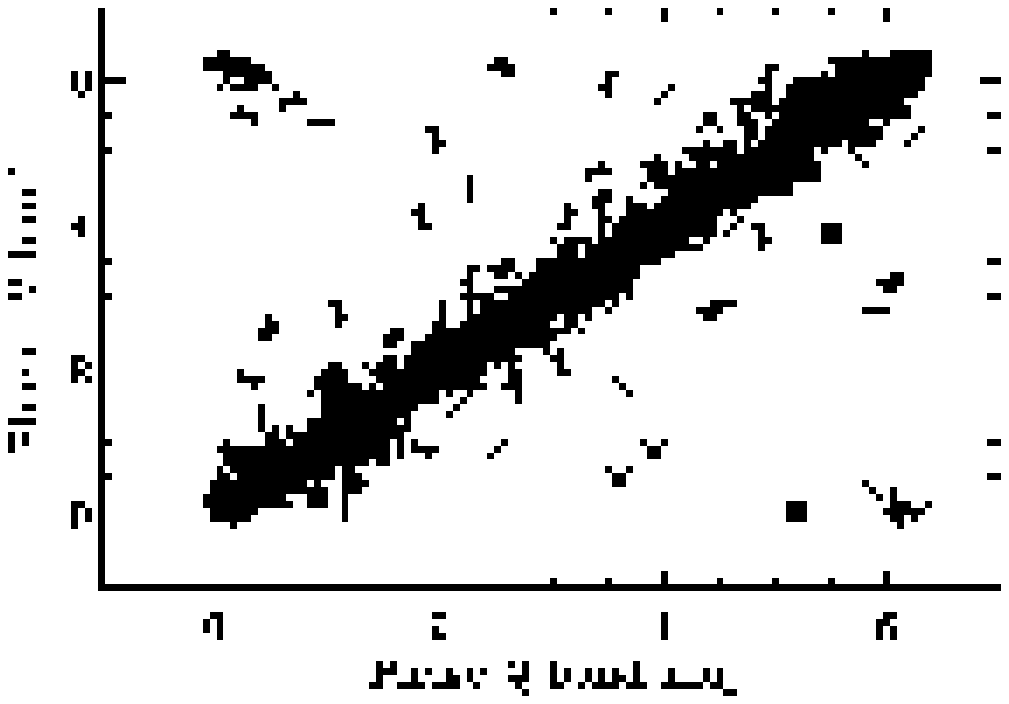}}}
\hbox{\hspace*{-0.5cm}
\centerline{\includegraphics[width=0.52\linewidth]{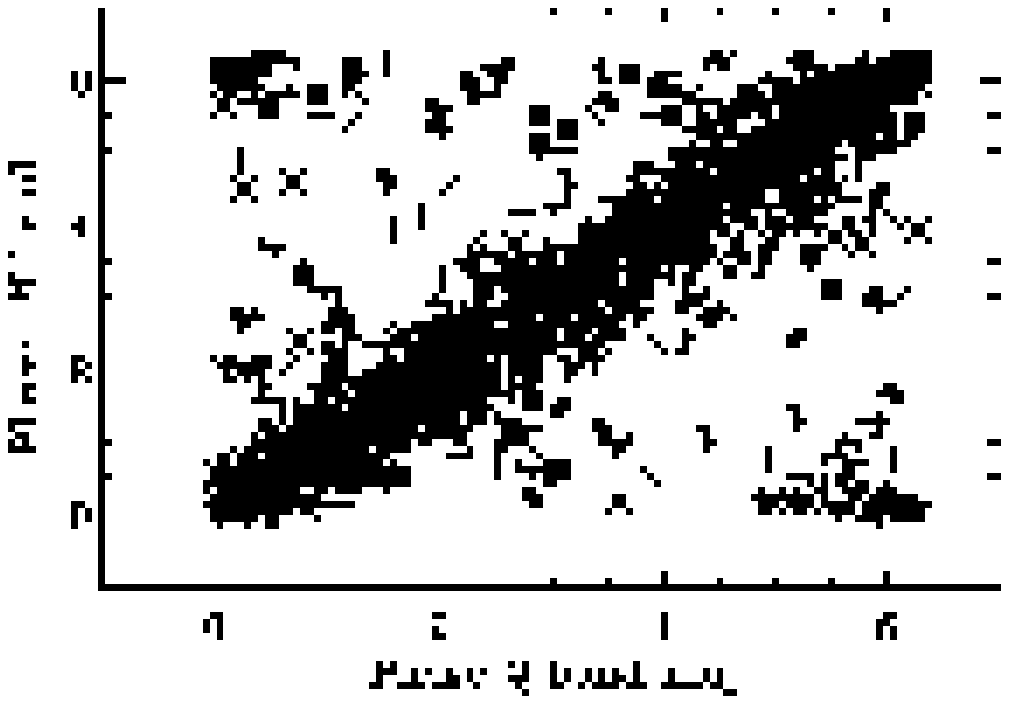}
\includegraphics[width=0.52\linewidth]{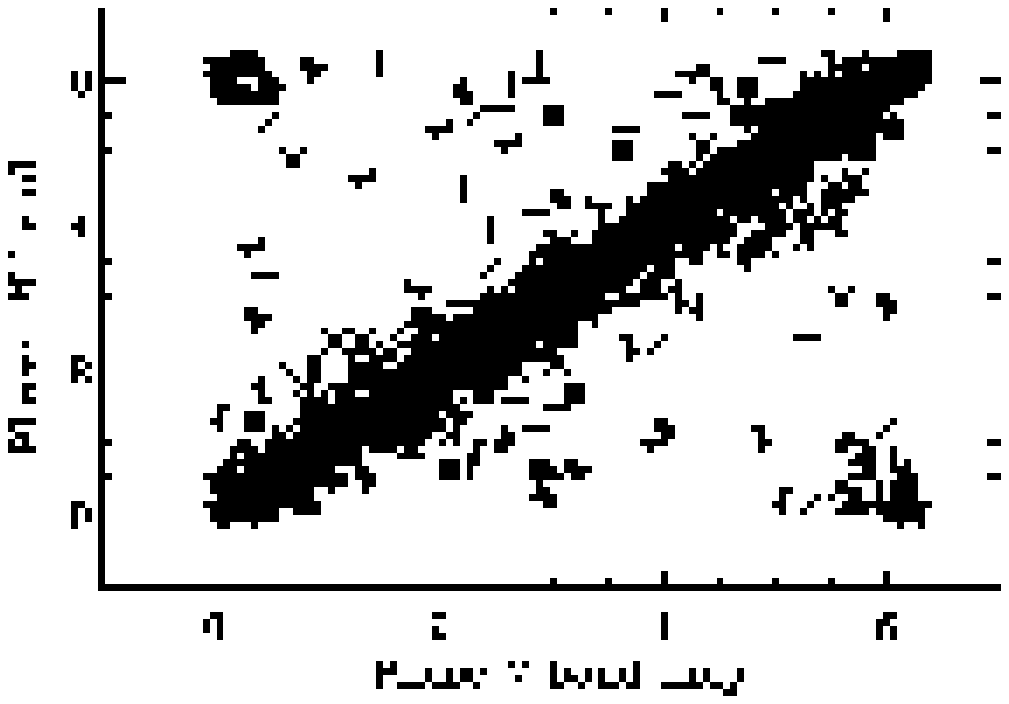}}}

\caption{The cross correlations of the phases between the maps of the
K--W bands.}
\label{xcorr} 
\end{apjemufigure}

As one can see from Eq.(\ref{eq7}) and (\ref{eq8}), the first term on the
right hand side of Eq.(\ref{eq7})
corresponds to the interference at the same multipoles $\ell,m$
between different foreground components, whereas the second and the
third correspond to between the noise--noise and between the foreground
component--noise interferences at the same frequency band, respectively.
We would like to point out that these interferences can be described in terms
of the auto-- and cross--correlations at the same channel.
However, because there is only one single realization for each frequency band
these correlations are not well--established. It would be difficult to
say, for example, that the instrumental noise does not correlate with
the foregrounds tails by using only the 5 channels and the corresponding 5
values for each $\ell,m$ harmonic.

In our phase analysis, therefore, we use some preliminary 
information about the possible values of the foreground and noise amplitudes
at different multipole range. That is, we discuss the properties
of the signals only at the range $\ell \le 50$, so that we can avoid
the beam shape and noise properties for all the K--W bands. This
multipole range is also the available range for the preliminary
foreground maps on the \wmap website.
Moreover, we can assume that the contribution from the
instrumental noise at this range is relatively small
in comparison with the
synchrotron, free-free and dust emissions (except for the K band). 
Generally speaking, we are investigating the properties of the CMB
signal at the \wmap K--W bands at the multipole range comparable with
the {\it COBE} range but slightly higher. At this range the
contamination of the foregrounds is the major part of the signal,
including the Galactic plane. 

\section{The ``Non-blind'' Phase--Cleaned Map (PCM)}
Note that we can neglect the instrumental
noise and the beam shape properties when we consider
multipole range $\ell \le 50$. It is obvious that if the foreground
signals are known, the pre-CMB signal can be
obtained from a simple subtraction of each band map by the
corresponding foreground map. We call such derived map the pre-CMB
signal because we assume that the foreground signals do not have the
required accuracy. As a result some residues from the foregrounds
propagate to the CMB signal. In reality, all properties of the
foregrounds and the pre-CMB signal come from the same combination of
the K--W maps and an uncertainties of the foreground and the CMB
signal estimation depend on each other. From a practical point of
view, therefore, any ``blind'' methods would be applicable in order to
find the correct phases of the CMB signal without any additional
assumption about the statistical properties of the foregrounds
\citep[for example]{te96,toh}. To make the first step for finding a
``blind'' method for the CMB reconstruction from the phases, we
consider first a ``non-blind'' method and will suggest a simple phase
filter, the modification of which allows us to reconstruct the pre-CMB phases
by a combination of the $\alm$ coefficients of the initial CMB + foregrounds
maps. This method generalizes the method from \citet{te96}, and the TOH
approach, but in our case we require the minimal variance in the phases
rather than in the derived map. 

We use all pairs of the \wmap maps for the 
reconstruction of the pre-CMB maps. Below we use the pair of maps $j=1,2$
(corresponding to the maps of the \wmap K and Ka bands) as
an example to derive the weighting coefficients, it can nevertheless be
applied to pairs such as $j=2,3$ (Ka,Q), $2,4$ (Ka,V) and $3,4$
(Q,V). We expect to obtain the derived map $M$ from the pairs of the \wmap
bands $M^{(1)}$ and $M^{(2)}$ maps, whose spherical harmonic coefficients are
$a^{(1)}_{\lm}$ and $a^{(2)}_{\lm}$, respectively. We have 
\begin{equation}
a^{M}_{\lm}= \sum_{j=1}^2 (-1)^{j+1} w^{(j)}_{\lm}a^{(j)}_{\lm} \equiv
\sum_{j=1}^2 (-1)^{j+1}\w^{(j)} \a^{(j)}, 
\label{eq9}
\end{equation}
with the normalization of the filter
\begin{equation}
\sum_{j=1,2} (-1)^{j+1}w^{(j)}_{\lm}=\I.
\end{equation}
Let us denote
$\w^{(j)}=|\w^{(j)}|\exp(- i\O^{(j)})$ where $|\w^{(j)}|$ and
$\O^{(j)}$ are the moduli and the phases of the weighting coefficients
for the $j$-th band, respectively. Because of the normalization
$\sum_j (-1)^{j+1}\w^{(j)}=\I$ the moduli and phases of the filter
$\w^{(j)}$ are not independent 
\begin{eqnarray}
|\w^{(1)}| & = & \frac{\sin \O^{(2)}}{\sin(\O^{(2)}-\O^{(1)})}, \nonumber \\
|\w^{(2)}| & = & \frac{\sin \O^{(1)}}{\sin(\O^{(2)}-\O^{(1)})},
\label{eq10}
\end{eqnarray}
and the resultant phases of the derived map $M$ are
\begin{eqnarray}
\lefteqn{\Ph^{M}=\tan^{-1}} \nonumber \\
& \left(\frac{\sum_j
  (-1)^{j+1}|\w^{(j)}||\G^{(j)}|\sin(\Ph^{(j)} - \O^{(j)}) + 
 |\a^{cmb}|\sin{\ks}}{\sum_j(-1)^{j+1}
  |\w^{(j)}||\G^{(j)}|\cos(\Ph^{(j)}- \O^{(j)}) +
  |\a^{cmb}|\cos{\ks}}\right),  
\label{eq11}
\end{eqnarray}
where $\G^{(j)}$ is the common signals excluding the CMB at the
$j$-th band and $\Ph^{(j)}$ is the phases of the $j$-th band. Thus,
if we prescribe that the reconstructed phase $\Ph^{M}$ from the
weighting coefficients is that of 
the pre-CMB signal (i.e. $\Ph^{M}=\ks$), we reach an exact solution
for the moduli and phases for the weighting coefficients
\begin{eqnarray}
|\w^{(1)}|& = &\frac{|\G^{(2)}|}{|\G^{(1)}-\G^{(2)}|}, \nonumber \\
|\w^{(2)}|& = & \frac{|\G^{(1)}|}{|\G^{(1)}-\G^{(2)}|}, \nonumber \\
\sin\O^{(1)}& = & |\w^{(2)}|\sin\Delta, \nonumber \\
\sin\O^{(2)}& = & |\w^{(1)}|\sin\Delta,
\label{eq12}
\end{eqnarray}
where $\Delta=\O^{(2)}-\O^{(1)}=\Ph^{(2)}-\Ph^{(1)}$.

Using the provided foreground maps from the \wmap database
and their 5 maps for the whole signal we are able to reconstruct the pre-CMB
signal using the filter Eq.(\ref{eq12}).
In Fig.\ref{precmb} we plot the reconstructed images of the pre-CMB
signal which contain some residues. For
drawing all maps we use $\ell \le 50$ only. The top panel is a simple
subtraction of the V band map by the foreground map in the V band. The
middle map was obtained using the filter Eq.(\ref{eq12}) and the
pair of the \wmap Q and V maps and the their
foreground maps for determination $\G^{(j)}$ and $\Delta$ in the filter
Eq.(\ref{eq12}). The bottom panel is the difference between the first
two panels.  

As one can see from Fig.~\ref{precmb} the derived pre-CMB maps look very
similar, while the contaminations from the bright point-like source
and diffuse residues exist notably along and perpendicular to the
Galactic plane, respectively. In order to compare our derived PCM and
the \wmap maps we plot the power spectrum of the 3 maps in Fig.\ref{Spectrum}, using a simple
pseudo-$\Cl$ estimator \citep{hivon},
\begin{equation}
\Cl=\frac{1}{2\ell+1}\sum_{m=-\ell}^{\ell}|\alm|^2.
\label{eq13}
\end{equation}

Note that our PCM has smaller power for the whole range of the multipoles
in comparison with that derived from the V band map. The flatness of 
the power spectrum at $\ell > 20$ clearly demonstrates the contamination of
the bright point-like source residues mainly concentrated along the
Galactic plane. 

\begin{apjemufigure}
\hbox{\hspace*{-0.2cm}
\centerline{\includegraphics[width=1.0\linewidth]{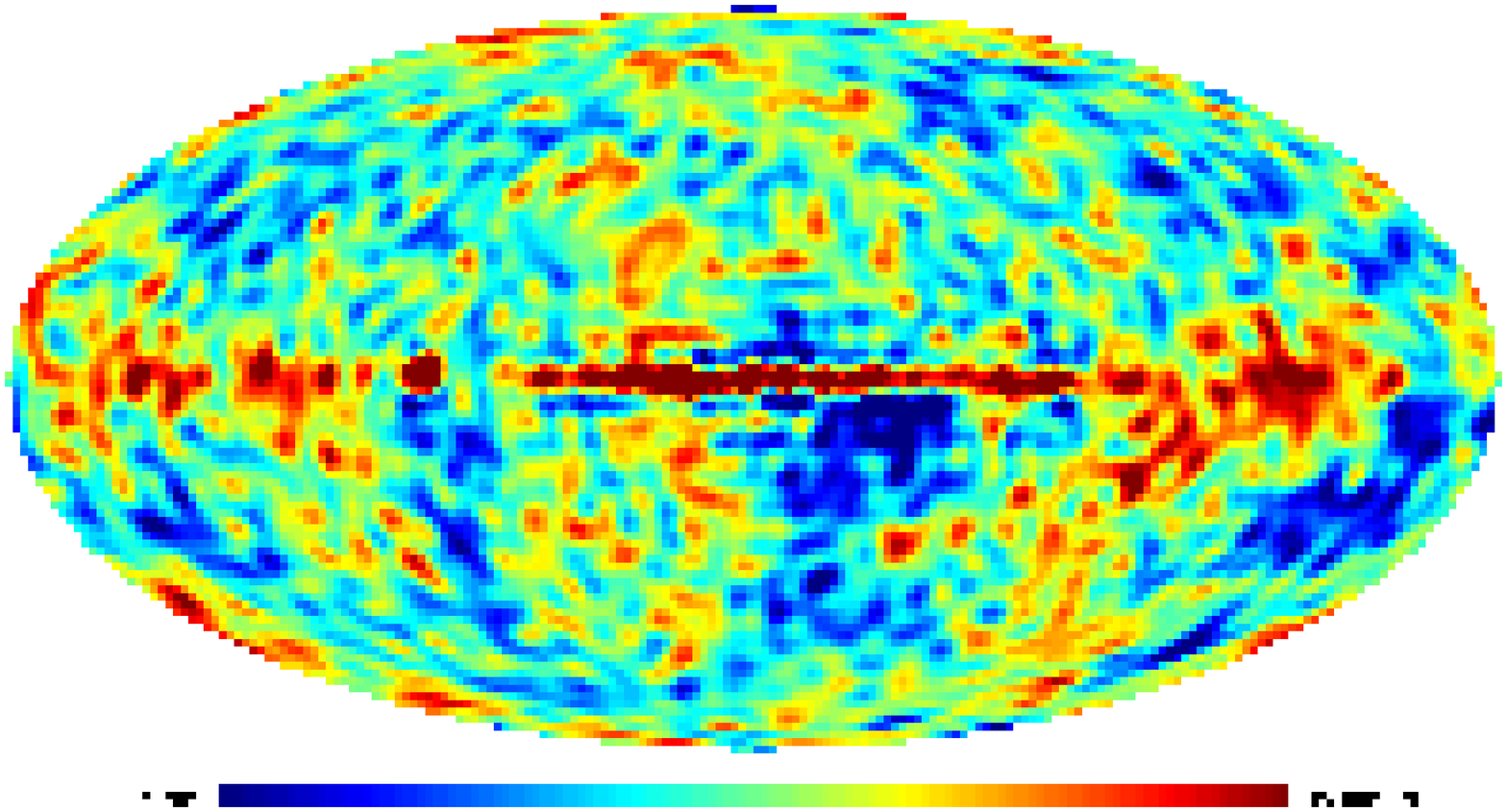}}}
\hbox{\hspace*{-0.2cm}
\centerline{\includegraphics[width=1.0\linewidth]{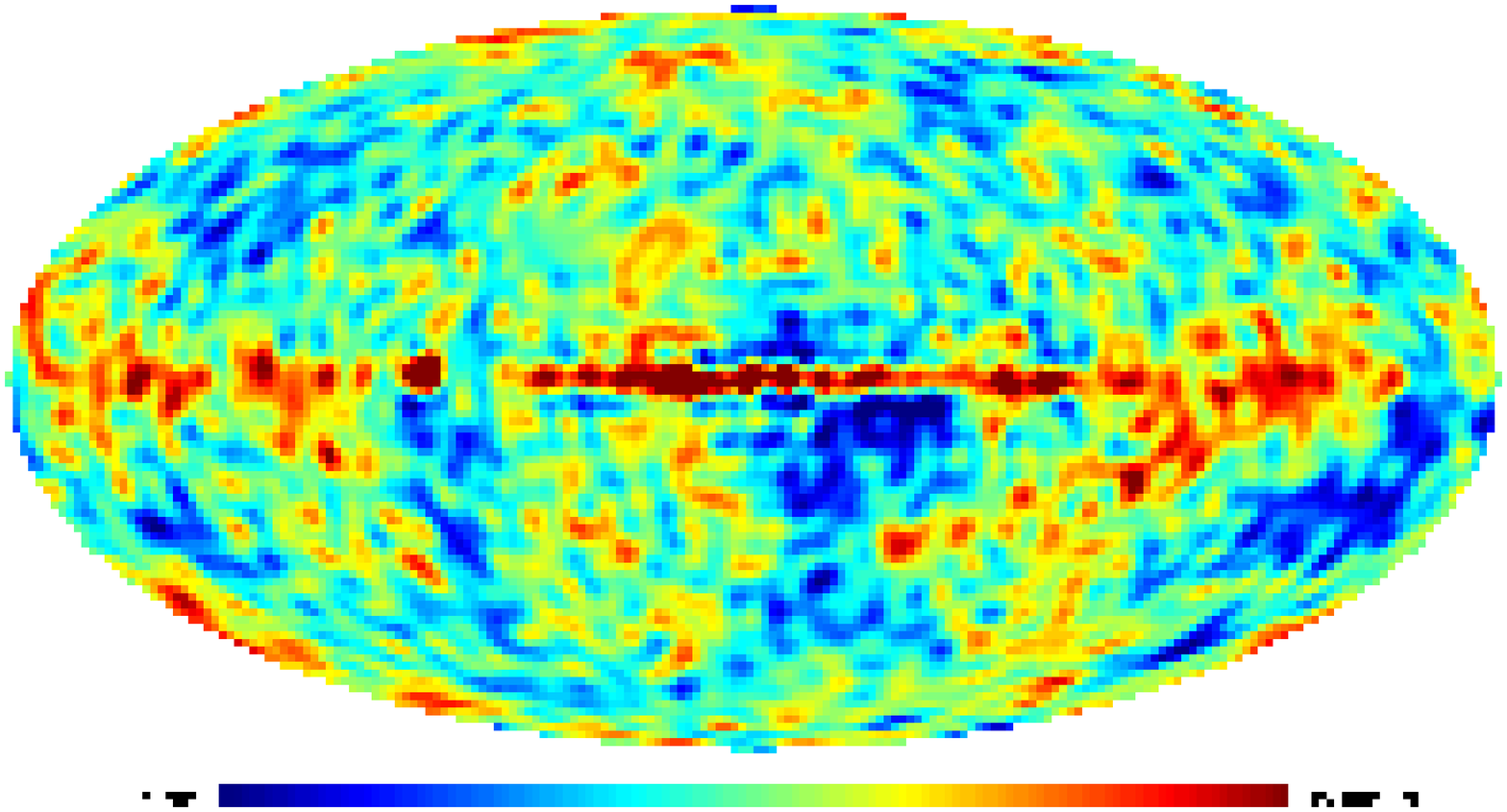}}}
\hbox{\hspace*{-0.2cm}
\centerline{\includegraphics[width=1.0\linewidth]{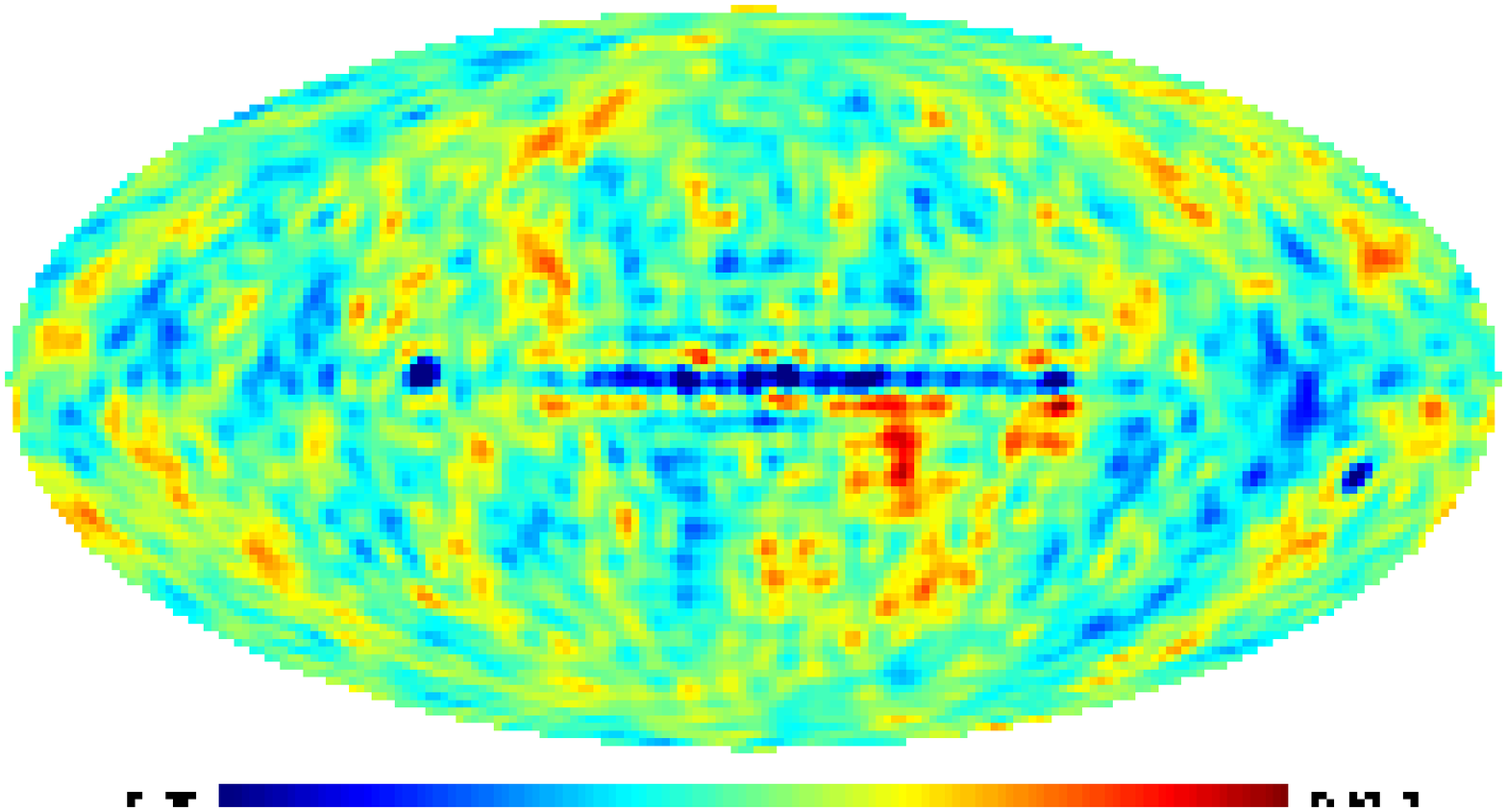}}} 
\caption{The maps for the derived pre-CMB signals.
Top panel is the map taken from a direct subtraction of
the foreground in V band by the \wmap total signal in V band. The
middle panel is the cleaned-up map obtained from the ``non-blind'' PCM
method from the \wmap Q and V bands, and the bottom the difference
between the first two panels.} 
\label{precmb} 
\end{apjemufigure}

\begin{apjemufigure}
\hbox{\hspace*{-0.1cm}
\centerline{\includegraphics[width=0.9\linewidth]{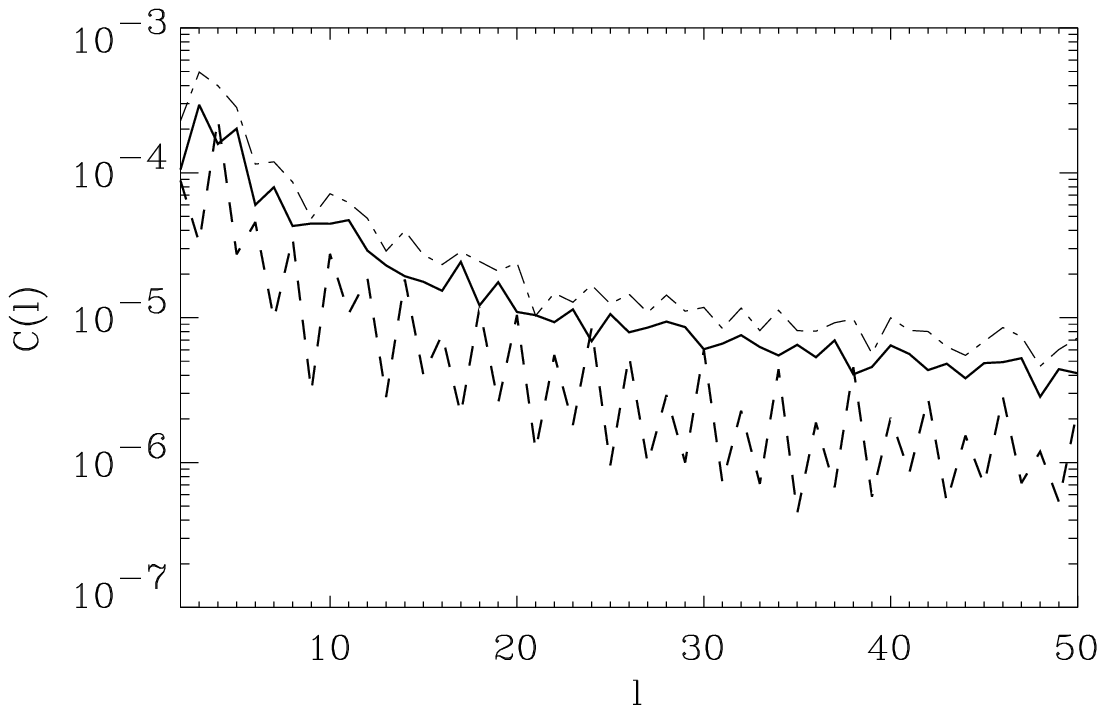}}}
\caption{The power spectrum for the derived from the \wmap V band map
(the dash dotted line), the PCM (the solid line) and the difference between
the two maps (the dash line).}
\label{Spectrum} 
\end{apjemufigure}

\section{The ``Blind'' reconstruction of the CMB}
In this section we describe the ``blind'' PCM method for separation of
the CMB signal from
the foregrounds using the K--W band signals at the harmonic range
$\ell \le 50$, avoiding again the beam convolution and the
instrumental noise in Eq.(\ref{eq5})-(\ref{eq8}).
The main assumption is that in the different frequency
bands the phases of the foregrounds do not
correlate with those of the true CMB signal. 
The result of separation improves if there are strong correlations between
the phases of the maps between the compared frequency bands
(see Fig.~\ref{xcorr}).

The basic idea of the ``blind'' PCM method is to generalize the
minimization scheme by TOH
and \citet{te96}, including the minimization of the
cross-correlations between derived
CMB signal and foregrounds phases. This ``blind'' method consists the
following four steps and does not need any galactic cut-offs and
dissection of the whole sky into disjoint regions.\\

{\bf a. We firstly group the following combinations 
of the \wmap maps: Ka--Q, Ka--V and Q--V.} They are grouped due to
highly correlated phases between the bands shown in
Fig.~\ref{xcorr}. We did not include the pair of the W and K bands in
the separation procedure because of the reasons mentioned in Appendix B.\\

{\bf b. For each pair of the maps (Ka--Q, Ka--V and Q--V) we
want to reach the pre-CMB map $\alm^M \equiv c_{\lm}$ derived from the
pair of the bands.} For Ka and Q ($j=2,3$), for example, we set
$M^{(2)} \equiv \alm^{(2)}$ and $M^{(3)} \equiv \alm^{(3)}$ and use
optimization coefficients ${\gam}^{(j)} \equiv \Gamma_{\ell}^{(j)}$,
similar to the TOH and \citet{te96}:
\begin{equation}
c_{\lm}=\sum_{j=2}^3
\Gamma_{\ell}^{(j)}\alm^{(j)} \equiv \sum_{j=2}^3{\gam}^{(j)}\a^{(j)} 
\label{eq14}
\end{equation}
The $\gam^{(j)}$ coefficients are subject to the following constraints
\begin{eqnarray}
\frac{\delta \sum_m|c_{\ell m}|^2} {\delta \gam^{(j)}}&  = & 0, \nonumber \\
  \sum_j\gam^{(j)} & = & \I, 
\label{eq15}
\end{eqnarray}
where $\delta/\delta \gam^{(j)}$ are functional derivatives.
The $\gam^{(j)}$ then have the following forms
\begin{eqnarray}
\gam^{(2)}& =\frac{\sum_m\left\{|\a^{(3)}|\left[\,|\a^{(3)}|-
    |\a^{(2)}|\cos(\Ph^{(3)}-\Ph^{(2)})\,\right]\right\}}
    {\sum_m|\a^{(2)}-\a^{(3)}|^2},  \nonumber \\
\gam^{(3)} &=\frac{\sum_m\left\{|\a^{(2)}|\left[\,|\a^{(2)}|-|\a^{(3)}|
\cos(\Ph^{(3)}-\Ph^{(2)})\,\right]\right\}} {\sum_m|\a^{(2)}-\a^{(3)}|^2}
\label{eq16}
\end{eqnarray}
where $\Ph^{(j)}$ are the phases of the total signal at
$M^{(j)}$ map.\footnote{We want to emphasize that one may think 
 it could be possible to omit the summation other $m$ in
Eq.(\ref{eq15}) and Eq.(\ref{eq16}) in order to optimize the phases at each
$\ell,m$ mode.  Then $\gam^{(j)}$ depend on $m$ too. The reasons why
we use the filter with the form Eq.(\ref{eq16}) are given in Appendix~A.}
After the pre-CMB separation, we obtain the pre-CMB signal $c_{\lm}$ and the
pre-foreground at each bands. \\

{\bf c. The next step is to find the filter  $\g^{(j)}$ to improve
the foreground separation using the phase variance minimization.} We
will use the filter $\g^{(j)}$ in Eq.(\ref{eq18}), which minimizes
the phase difference between reconstructed and the CMB phases. We
assume that $\g^{(j)}$ does not possess any imaginary part and depends
only on $\ell$ but not on $m$.  The theoretical basis of our choice of
the form of the filter is for minimization of the weighting phase variance
\begin{equation}
\mbox{\~V} =\frac{1}{2 \pi(2\ell+1)}\sum\limits_m \frac{|a_{\ell
    m}^{cmb}|^2}{\Cl} (\Ph-\ks)^2 \rightarrow \min
\label{eq17}
\end{equation}
This condition is equivalent to minimization of the
error $\varepsilon=(2\ell+1)^{-1}\sum_m |\alm^{M} - a_{\ell m}^{cmb}|^2$
between reconstructed ($\alm^{M}$) and the true ($a_{\ell m}^{cmb}$) CMB
signals\footnote{See Appendix B for details}.
 Using foregrounds estimator $\G^{i}=\a^{i}-\a^{M}$ from the step {\bf b}, 
we obtain the following filter $\g^{(j)}$ instead of the
filter (\ref{eq12}).

\begin{eqnarray}
\g^{(2)}& = \frac{\sum_m|\G^{(3)}|\left[\,|\G^{(3)}|-
|\G^{(2)}|\cos(\Ph^{(3),f}-\Ph^{(2),f})\,\right]}{\sum_m|\G^{(2)}-\G^{(3)}|^2}
&\nonumber\\ 
\g^{(3)}& = \frac{\sum_m|\G^{(2)}|\left[\,|\G^{(2)}|-|\G^{(3)}|
\cos(\Ph^{(3),f}-\Ph^{(2),f})\,\right]}{\sum_m|\G^{(2)}-\G^{(3)}|^2}
\label{eq18}
\end{eqnarray}
where $\Ph^{(j),f}$ are the phases of the foregrounds at the $j$-th
band. This filter provides a new estimation of the foregrounds and the
pre-CMB signals which we shall use for the next step of the subsequent
iterations by the filter Eq.(\ref{eq18}). So, at each iteration we insert
the foreground moduli and phases obtained from the previous iteration
to the same form of the filter Eq.(\ref{eq18}). In general, 3 such iterations are enough to produce a convergent map.\\

{\bf d. We use a MIN-MAX filter to further clean up the residues.}
After the step ${\bf c.}$ we have three cleaned-up maps from the \wmap
map pairs:  Q--V, Ka--V and Ka--Q bands. These 3 maps correspond to
the minima of the phase difference (see Fig.\ref{xcorr}). They
nevertheless still contain some of the foreground residues, notably
along the Galactic plane. In order to detect and minimize such a
residues we used a MIN-MAX filter on the same pixel among the 3 cleaned-up
maps. The definition of the MIN-MAX filter is the
following. Using maps with the same pixelization scheme, we can
designate all the pixels in the same way and compare simultaneously at
the same pixel $p$ these 3 cleaned-up maps $\Delta T^{(i)}_p$ by
pairing them up, where
the index corresponds to the 3 cleaned-up maps. The MIN filter
$L_{\min}$ allows us to find the signal which has a minimal
(absolute) amplitude at the pixel $p$, i.e. $L_{\min}(\{\Delta
T^{(i)}_p \}) \rightarrow \Delta T^{\min}_p$, where $|\Delta
T^{\min}_p |=\min\{|\Delta T^{(i)}_p|\}$.   
 
In other words, we take the smallest absolute amplitude at
the pixel $p$ from the cleaned-up maps and attach its sign for that
pixel (or even more concisely, the value which is the closest to zero) .  

The reason for such a filter is clear. The signal at each pixel is a
combination of the underlying CMB signal and a small residue from the
foreground cleaning. If the derived CMB signal correlates little with the
foregrounds, it is natural to expect that all deviations of
the  $\Delta T^{(i)}_p$ at $p$ of the different maps are caused by the
residues but not by the CMB. In addition to the $L_{\min}$ filter we
can define a MAX filter, $L_{\max}(\{\Delta T^{(i)}_p\}) \rightarrow
\Delta T^{\max}_p $ which is the 
farthest value from zero between the maps at pixel $p$. This MAX filter
reproduces the most possible contribution from the foreground residues
at each pixel. Thus, we find our final PCM, the true CMB map according to
the criterion that $\Delta T^{\max}_p - \Delta T^{\min}_p$ map has
the mininum power. We apply our MIN-MAX filter on Q--V,
Ka--V and Ka--Q cleaned-up maps and obtain the PCM signal from Q--V,
Ka--V maps combination.

At the end of this Section we note the following comment. In
Appendix~C we show
analytically that if the phase $\ks$ of some $\alm$ of the CMB signal is
close to $\pi/2$ or $3\pi/2$ the error of the phase reconstruction is
very high.
Practically these phases can not be reconstructed. We call this problem
the ``$\pi/2$'' problem. Our numerical tests (see section 5) confirm
this conclusion.

\begin{apjemufigure}
\hbox{\hspace*{-0.2cm}
\centerline{\includegraphics[width=1.\linewidth]{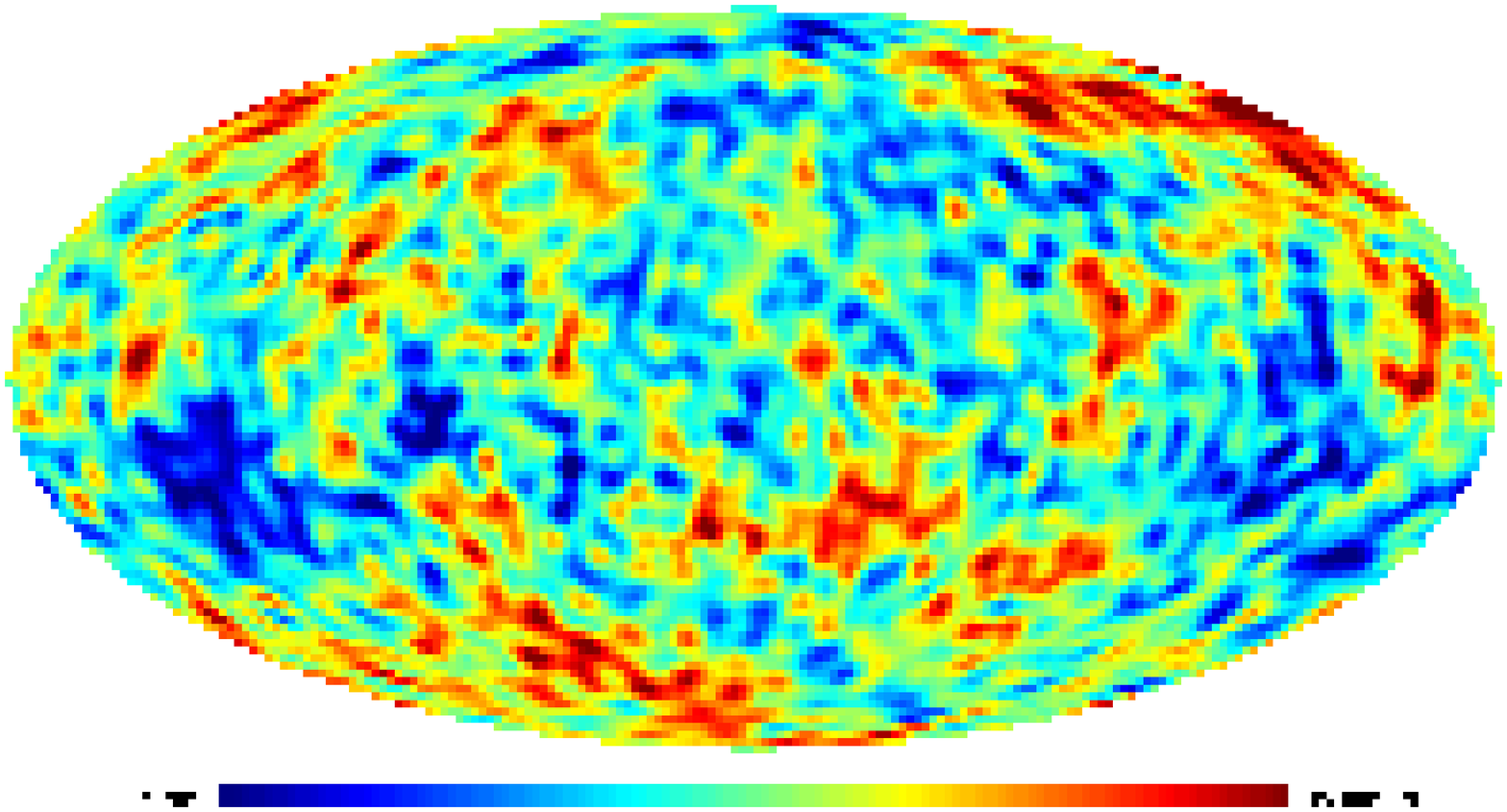}}}
\hbox{\hspace*{-0.2cm}
\centerline{\includegraphics[width=1.\linewidth]{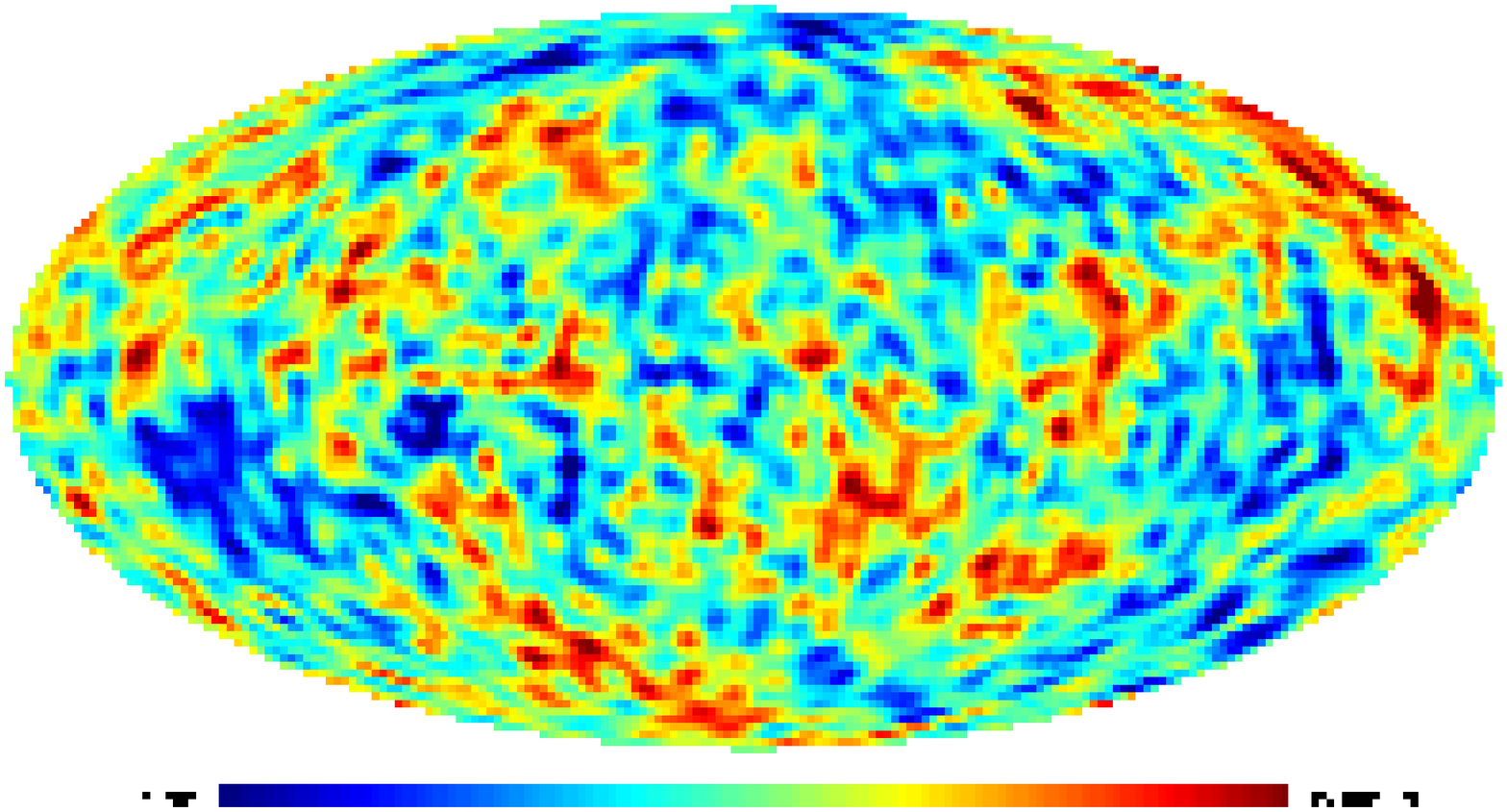}}}
\hbox{\hspace*{-0.2cm}
\centerline{\includegraphics[width=1.\linewidth]{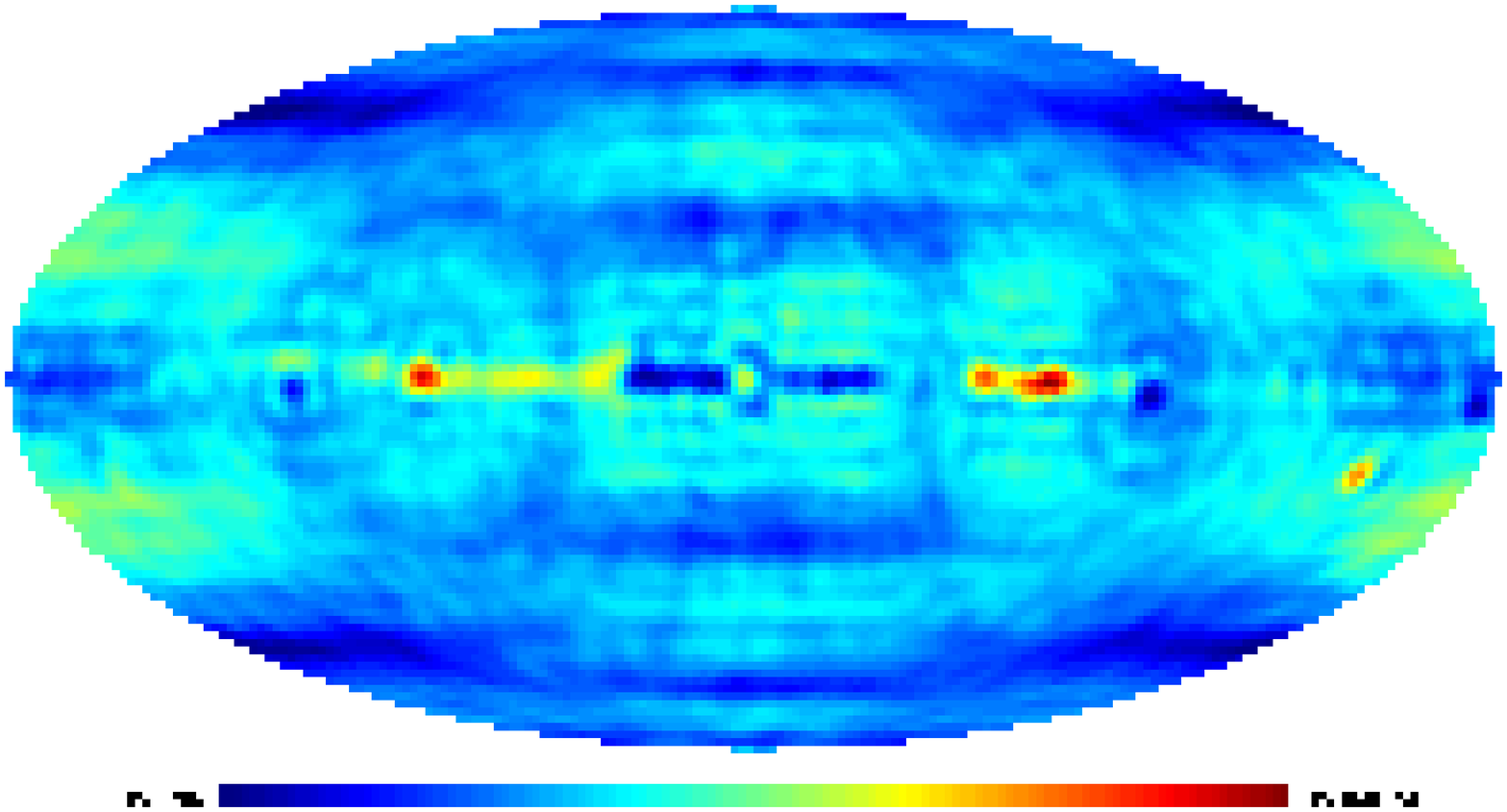}}}
\caption{The reconstructed PCM from the \wm simulator. Top panel is
the simulated CMB map and the middle is the reconstructed PCM from the
two pairs of simulated bands, Q--V and Ka--V. The bottom
panel is the difference between the two.}
\label{simulator} 
\end{apjemufigure}

\section{The numerical test}
To numerically test the ``blind'' PCM method we use a \wm
simulator to mimic all the properties of the \wmap foregrounds at K-W
bands. We again exclude the instrumental noise and the beam shape
properties, which are insignificant for the multipole range $\ell \le 50$.
We take the best-fit \wmap $\Lambda$CDM power spectrum to produce a
simulated CMB sky with random phases. Such CMB sky is truly a Gaussian
random realization. The morphology of this simulated
CMB map is obviously different from that of the \wmap ILC and the TOH
maps (since it has a different set of random phases).
For our \wm simulator we adopt all the foreground maps for each
frequency band from the \wmap website. 
We also confirm that the simulated CMB signal does not have any
cross-correlation with the foregrounds and it has well-defined
statistical properties (the power spectrum and random phases).

\begin{apjemufigure}
\hbox{\hspace*{-0.1cm}
\centerline{\includegraphics[width=0.9\linewidth]{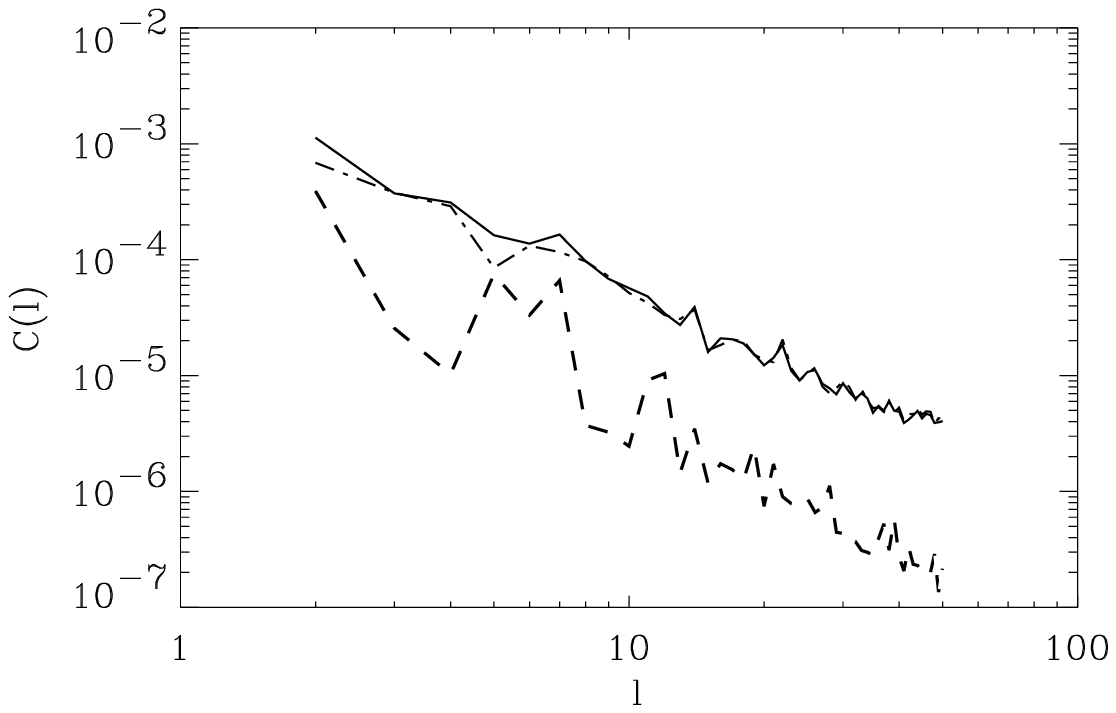}}}
\caption{The reconstructed power spectrum of the \wm simulator from
the ``blind'' PCM method. The
solid line is the power spectrum of the simulated CMB signal and the
dash-dot line is that of the reconstructed PCM. Note that both curves
almost overlap except for $\ell=2$, 5 and 7. The dashed line is the
difference of these two. These 3 power spectra correspond to the 3
panels in Fig.\ref{simulator}. The reconstructed PCM signal is produced from
two pairs of the simulated channels, Q--V and Ka--V.} \label{simulatorpw} 
\end{apjemufigure}

\begin{apjemufigure}
\hbox{\hspace*{-0.1cm}
\centerline{\includegraphics[width=0.9\linewidth]{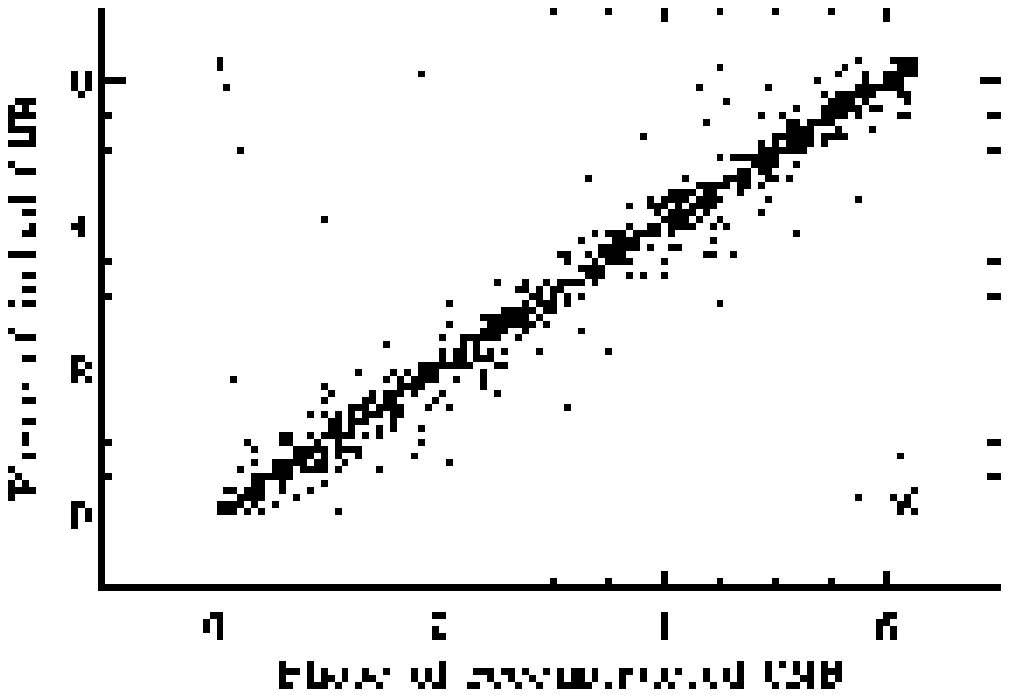}}}
\caption{The cross correlation of the phases between the simulated
(random-phase) CMB and the reconstructed CMB (PCM) signal. The solid
 diagonal line indicates an asymptotic when the simulated and reconstructed
phases are identical.}
\label{pcmxcorr} 
\end{apjemufigure}

In Fig.\ref{simulator} we show the result of the CMB reconstruction after
10 iterations from the ``blind'' PCM method.
As we mention above the third and all subsequent iterations
practically produce a convergent cleaned-up map.
We take two pairs of simulated channels, Q--V and Ka--V, to
reconstruct the signal. The result of reconstruction for both pairs
was practically the same, and the MIN-MAX filter is used
to produce the resultant reconstructed map shown in Fig.\ref{simulator}.
In Fig.\ref{simulatorpw} we plot the power spectra for the simulated 
and the reconstructed CMB (PCM) signals as a function of $\ell$. Both
curves almost overlap except for $\ell=2$, 5 and 7.

In Fig.\ref{pcmxcorr} we plot the cross-correlation of the phases
between the simulated CMB signal and the reconstructed CMB (PCM).
As one can see from Fig.\,\ref{pcmxcorr} for reconstructed CMB signal
the quadrupole component has smaller power than the simulated CMB.
The reason for that is that the peculiar phases of the quadrupole component
$Q_{20}$, $Q_{21}$ and $Q_{22}$. Accidentally, in our
simulated realization of the CMB (see Fig.\ref{simulator})
the phases of the component $Q_{20}$ and  $Q_{21}$ are zero whereas
for the $Q_{22}$ component the corresponding phase is close to
$3\pi/2$ with only 10\% deviation. That is why the reconstructed phase
of $Q_{22}$  component has $\ks_{22}=6.227$ radians. 

In Fig.\ref{simulatorxcorr} we plot the cross correlation of the
phases between the reconstructed CMB signal (PCM) and the simulator
foregrounds at all 5 simulated bands. The scattering of points in all 5 panels
indicates that the phases of the PCM display no serious correlations
with those of the foregrounds. The \wm simulator therefore confirms the
usefulness of our ``blind'' PCM method for the CMB extraction without
any assumptions about the statistical properties of the foregrounds.

\begin{apjemufigure}
\hbox{\hspace*{-0.5cm}
\centerline{\includegraphics[width=0.52\linewidth]{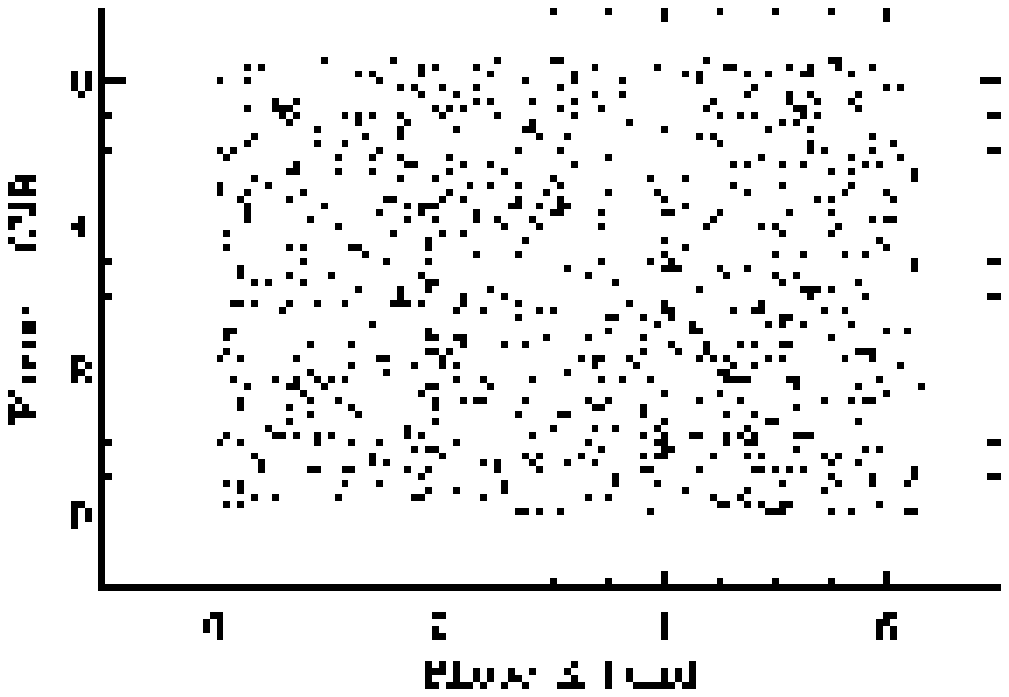}
\includegraphics[width=0.52\linewidth]{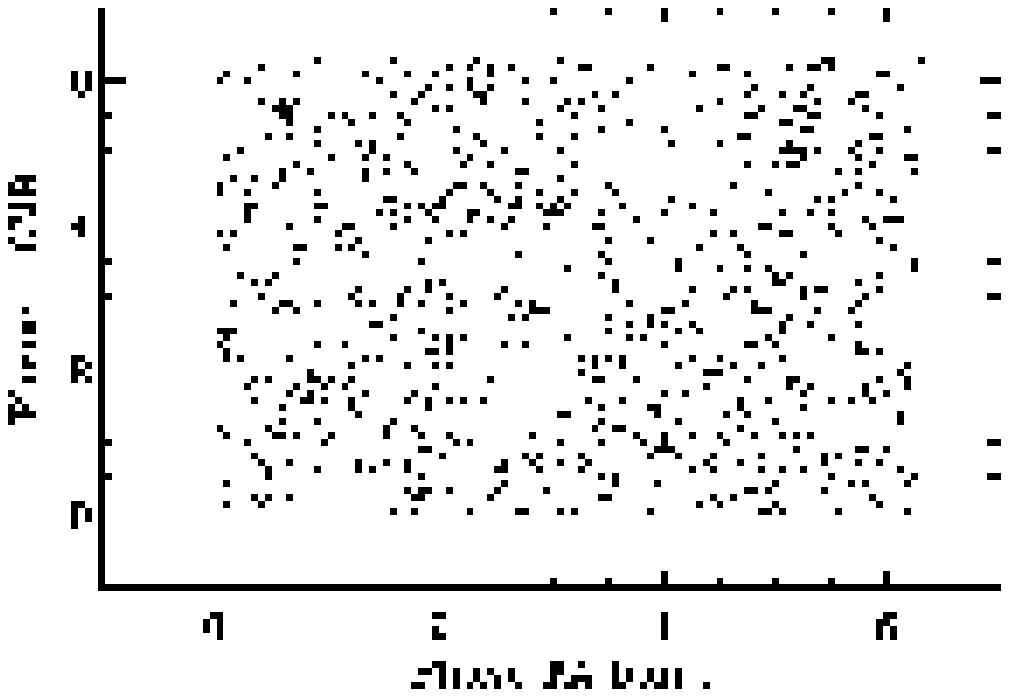}}}
\hbox{\hspace*{-0.5cm}
\centerline{\includegraphics[width=0.52\linewidth]{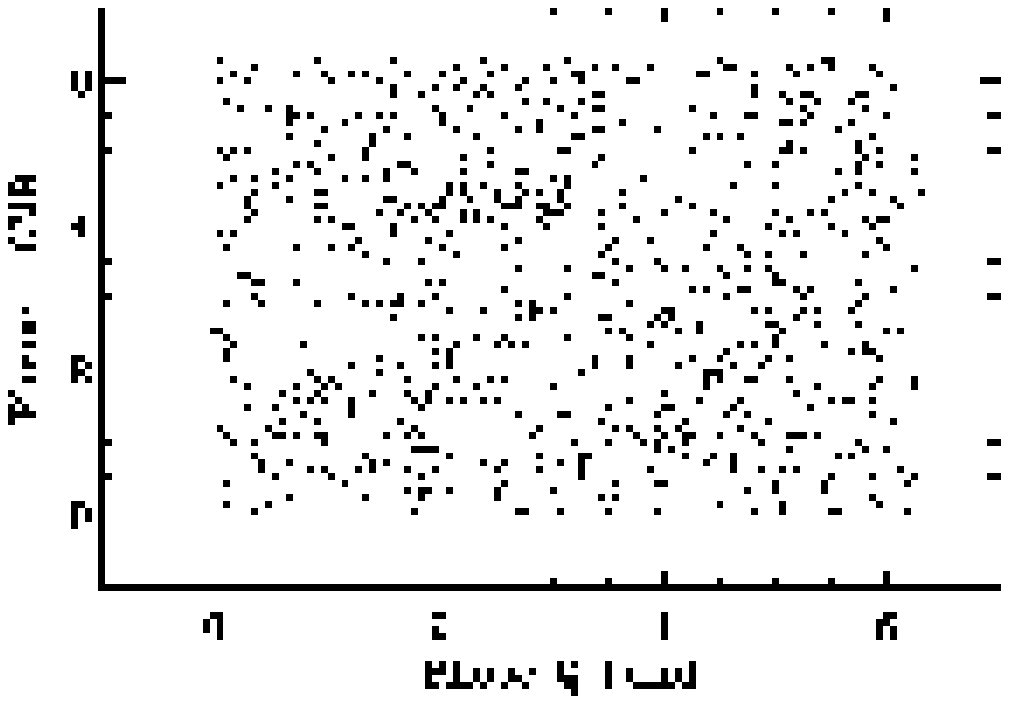}
\includegraphics[width=0.52\linewidth]{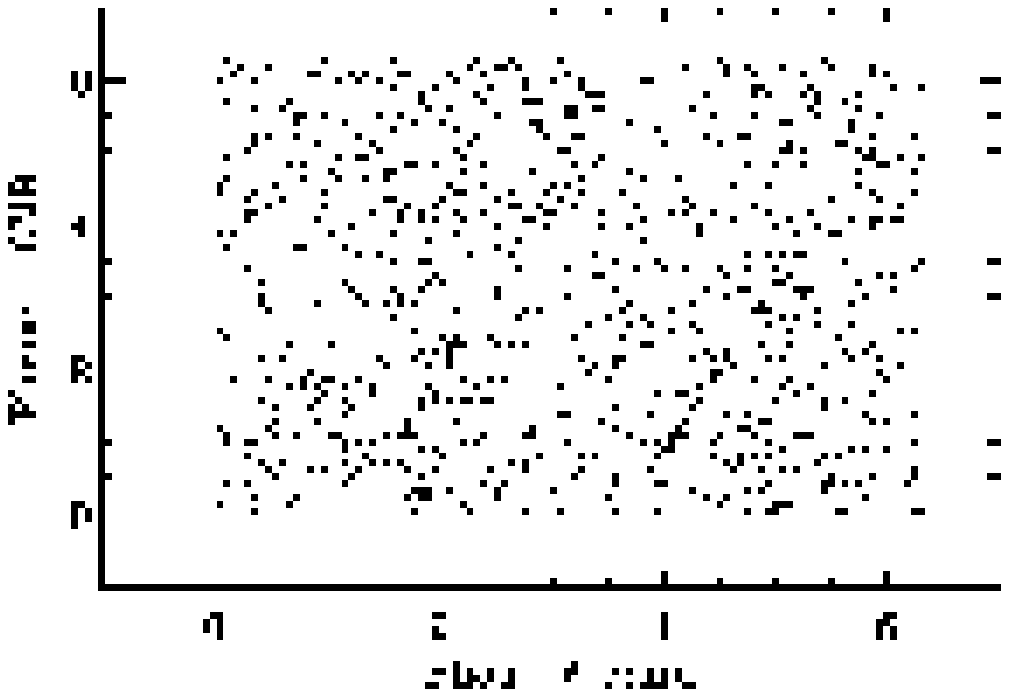}}}
\hbox{\hspace*{-0.5cm}
\centerline{\includegraphics[width=0.52\linewidth]{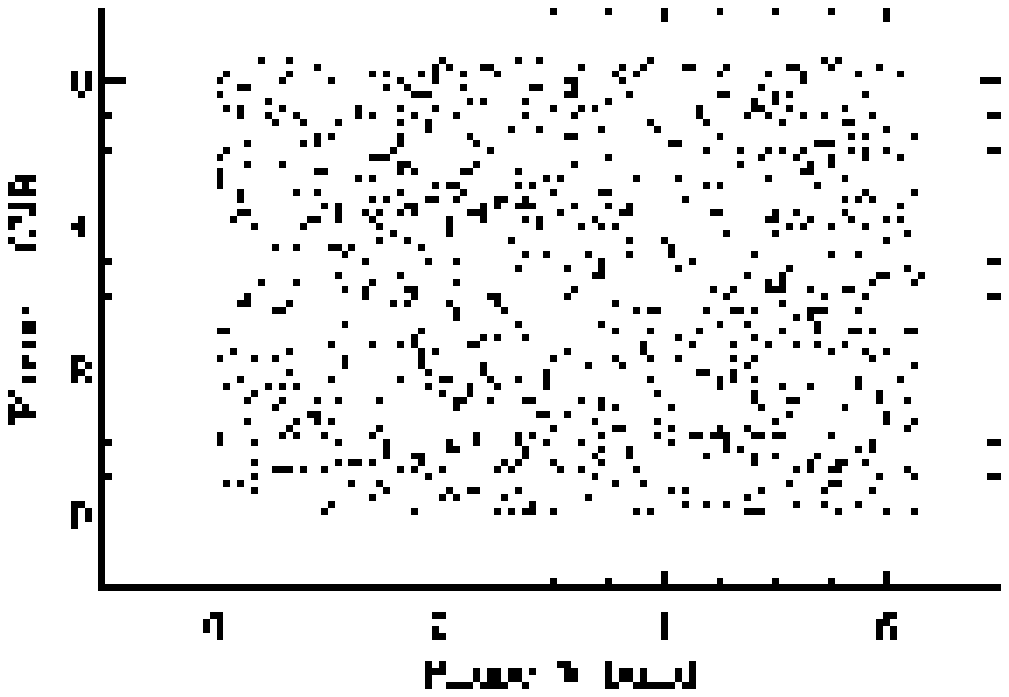}}}
\caption{The cross correlations of the
phases between the reconstructed CMB signal (PCM) and the simulator
foregrounds of all 5 simulated bands. The scattering of points in all 5 panels
indicates that the phases of the PCM display no correlations with those
of the foregrounds.}
\label{simulatorxcorr}
\end{apjemufigure}

\section{The PCM derived from the 1-year \wmap data}
For the reconstruction of the CMB signal at $\ell \le 50$ from the
K--W bands of the \wmap data we use the three pairs of the maps:
Ka--Q, Q--V and Ka--V. We repeat the steps ${\bf b}$ and ${\bf c}$ of
the ``blind'' PCM method : each of them has been convolved by the weighting
coefficients, Eq.(\ref{eq16}) and then by the coefficients
Eq.(\ref{eq18}) using only 2 iterations. All the next iterations do
not change significantly the $\alm$ coefficients (the corresponding
error being less than $10^{-3}\%$). The results are present in
Fig.\ref{cleanedup}, in which all the 3 cleaned-up maps have pronounced
similarity in morphology, albeit with some residues from the galactic
foregrounds.

\begin{apjemufigure} 
\hbox{\hspace*{-0.2cm}
\centerline{\includegraphics[width=1.0\linewidth]{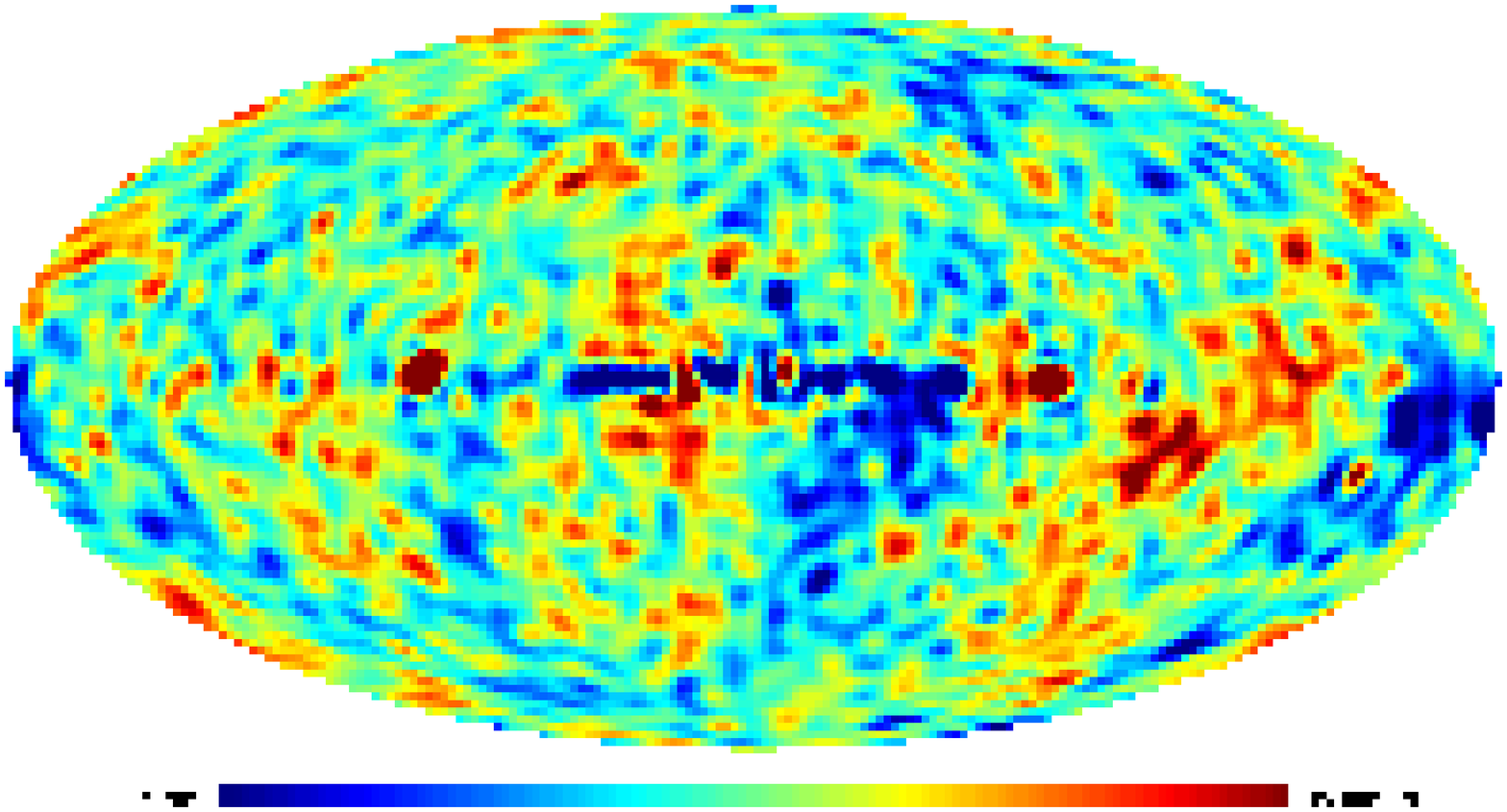}}}
\hbox{\hspace*{-0.2cm}
\centerline{\includegraphics[width=1.0\linewidth]{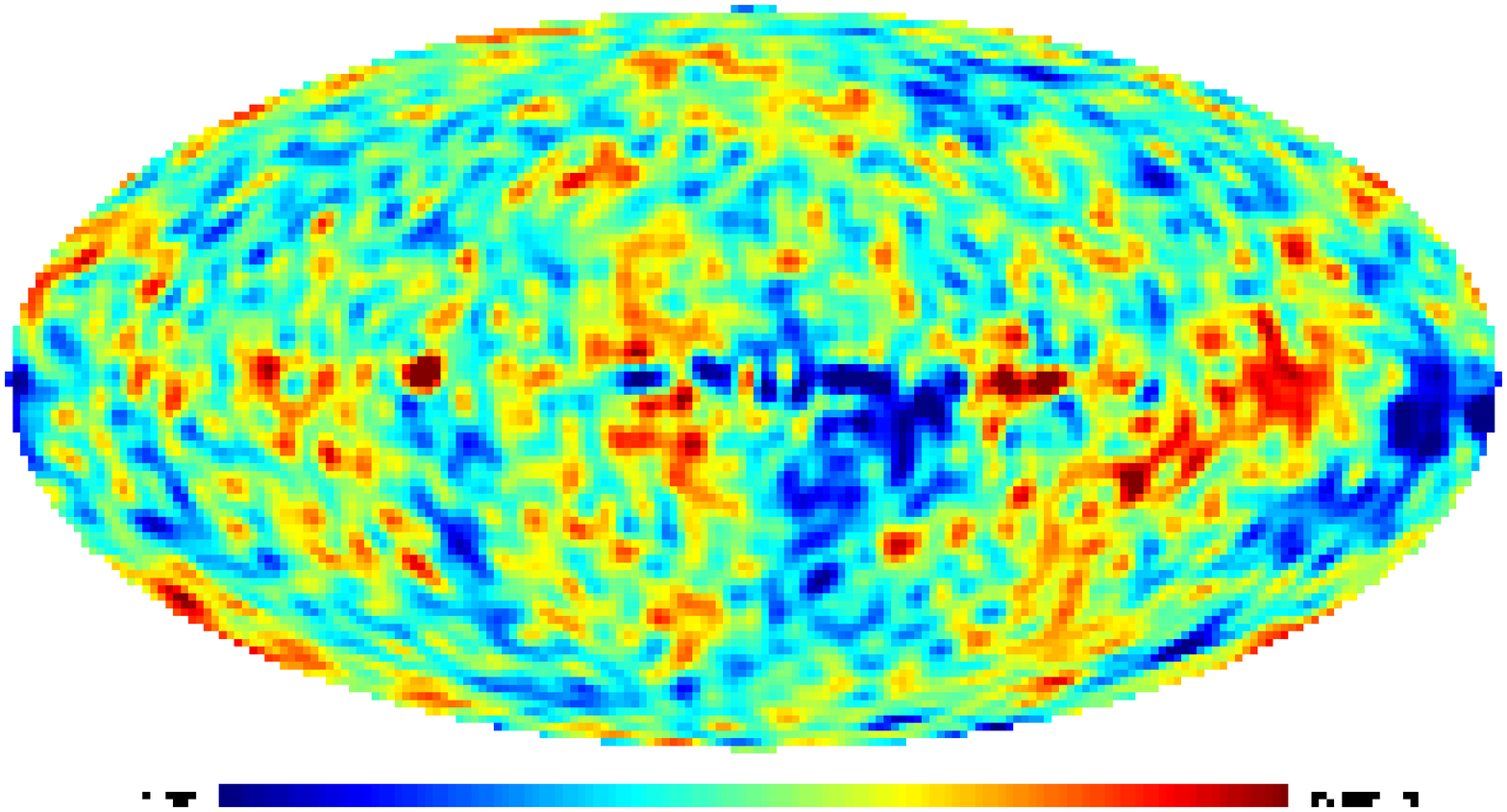}}}
\hbox{\hspace*{-0.2cm}
\centerline{\includegraphics[width=1.0\linewidth]{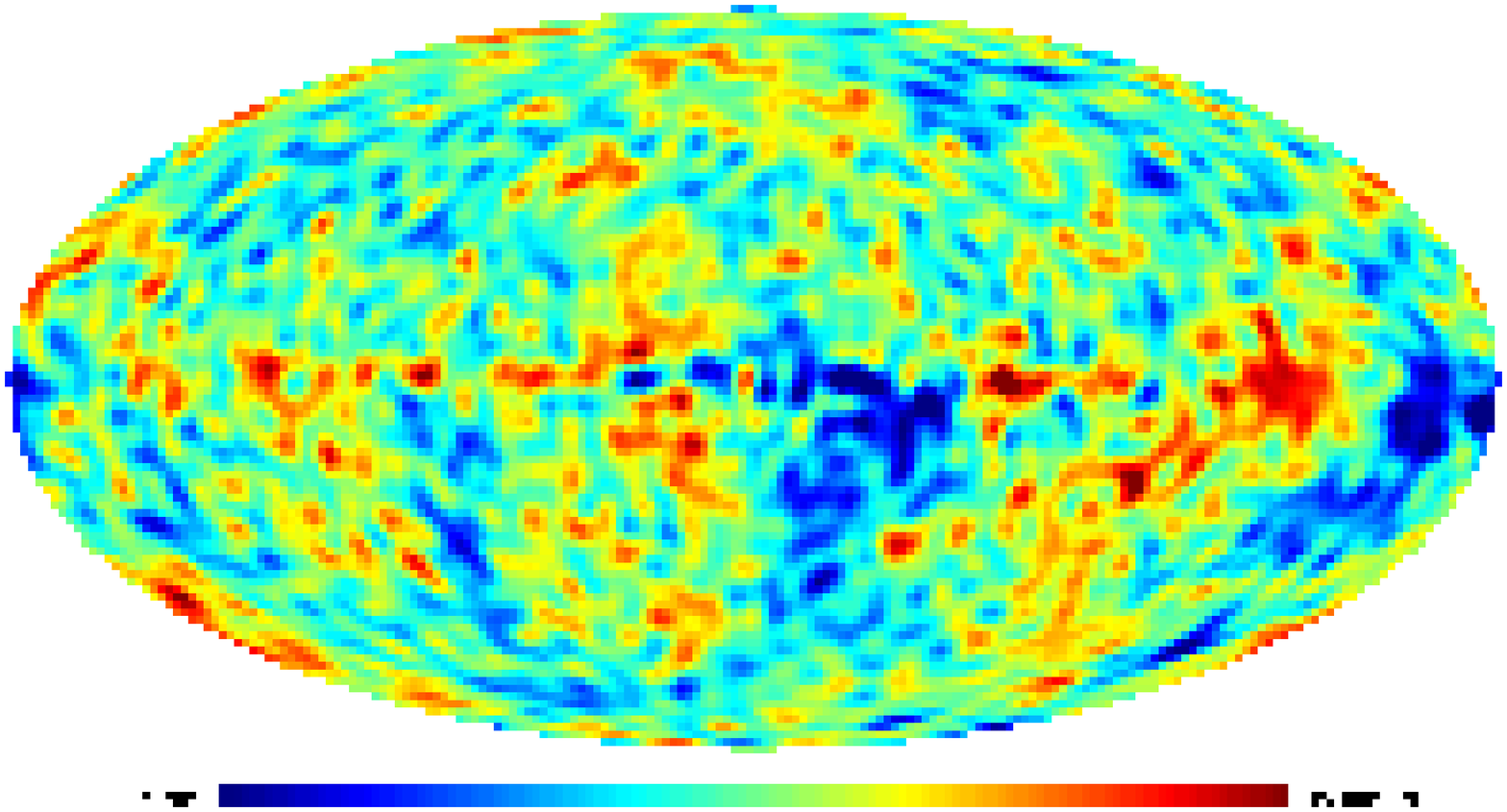}}}
\caption{The cleaned-up maps after the steps ${\bf b}$ and ${\bf c}$ from
 the ``blind'' PCM method on the 3 pairs of \wmap maps. The top
 panel is the cleaned-up map from the Ka--Q maps, the middle the
 cleaned-up map from the Ka--V maps, and the bottom from the Q--V
 maps.}
\label{cleanedup} 
\end{apjemufigure}

In Fig.\ref{minmax} we show the result after the step ${\bf d}$, the
MIN-MAX filtering. The top panel is our CMB signal (the PCM), the
middle is the \wmap ILC map for comparison and the bottom is the
difference of the two.

In Fig.\ref{pcmpw} we plot the CMB angular power spectra from different
methods: the thick solid line is the PCM, the dashed line the \wmap
ILC map, the dash-dot line the TOH foreground-cleaned map, and the
thin solid line the TOH Wiener-filtered map. The PCM power spectrum
is smaller then ILC power spectrum at the whole multipole range $\ell
\le 50$ and it reproduces a similar power spectrum as the TOH
Winer-filtered map. 

In Fig.\ref{pcmwmapxcorr} we show the cross correlation of the
phases between the PCM and the foreground maps of the \wmap
bands. Again the scattering of the points shows that the phases of the
PCM practically have no correlations with those of the foregrounds.

\begin{apjemufigure} 
\hbox{\hspace*{-0.2cm}
\centerline{\includegraphics[width=1.0\linewidth]{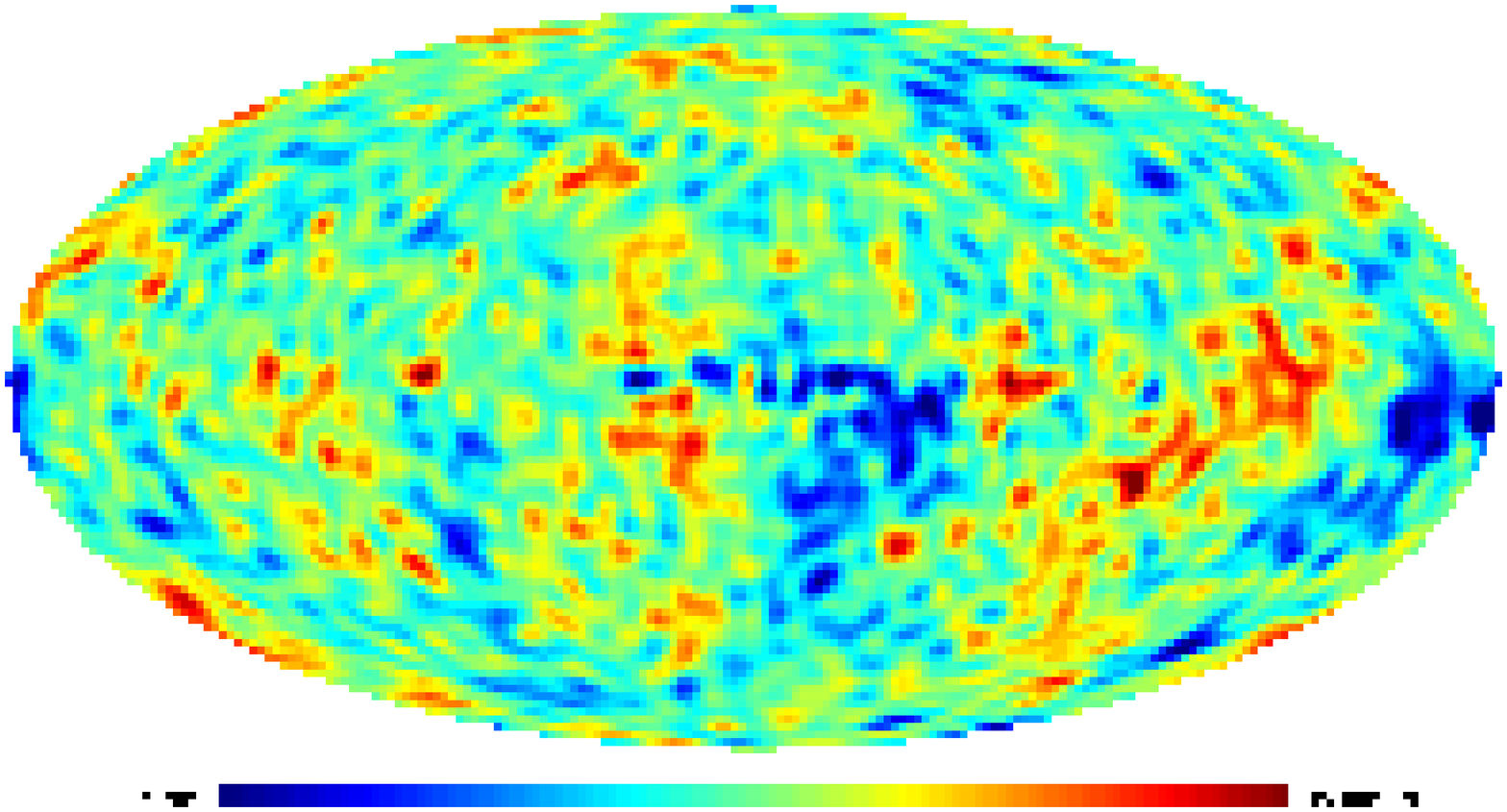}}}
\hbox{\hspace*{-0.2cm}
\centerline{\includegraphics[width=1.0\linewidth]{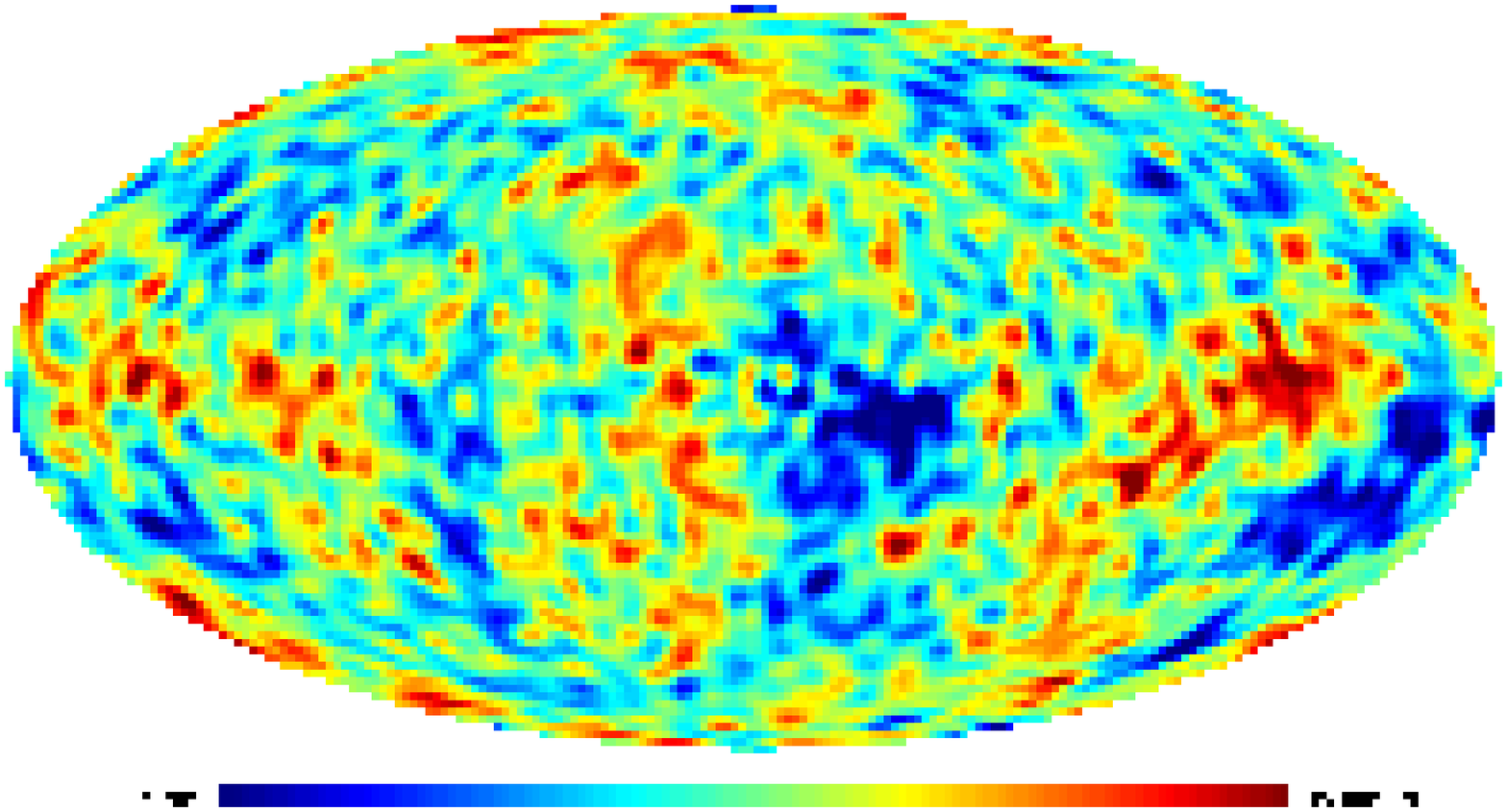}}}
\hbox{\hspace*{-0.2cm}
\centerline{\includegraphics[width=1.0\linewidth]{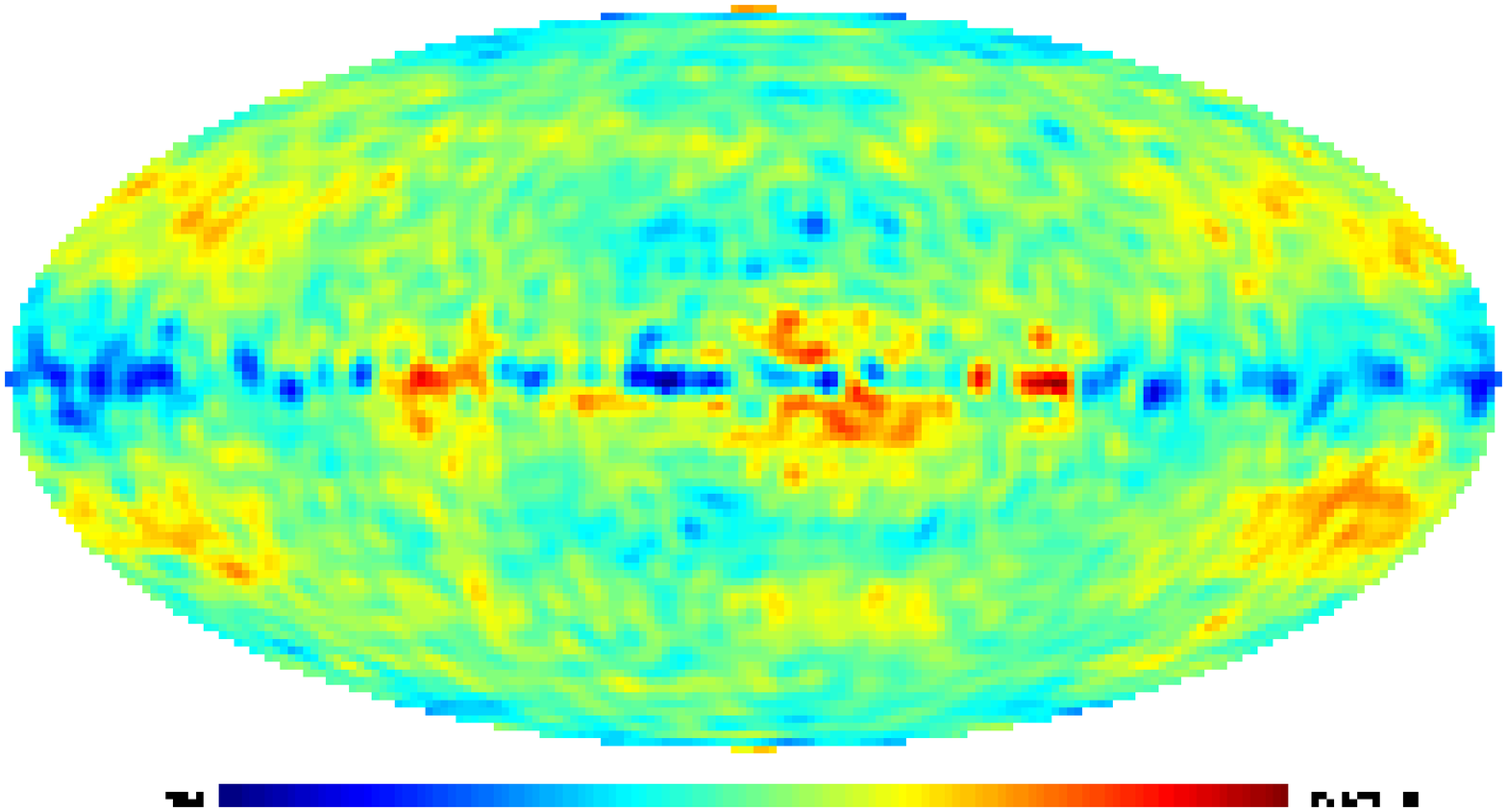}}}
\caption{The CMB map (the PCM) after the step ${\bf
d}$ of the ``blind'' PCM method (top panel). The MIN-MAX filter is
performed on the 3 panels of Fig.\ref{cleanedup}. For comparison, the
middle is the \wmap ILC map and the bottom is the difference of the
two. }
\label{minmax} 
\end{apjemufigure}

\begin{apjemufigure} 
\hbox{\hspace*{-0.1cm}
\centerline{\includegraphics[width=0.9\linewidth]{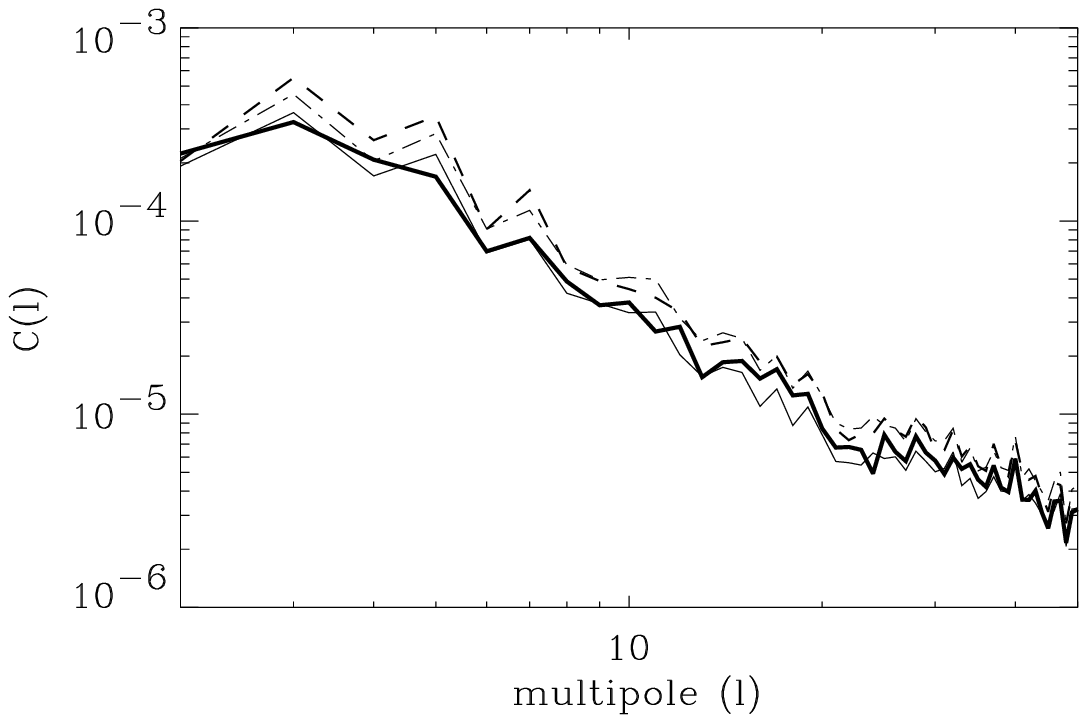}}}
\caption{The comparison of the CMB angular power spectra from different
methods. The power spectrum of the PCM is shown with the thick
solid line, that of the \wmap ILC map is shown with the dashed
line, the TOH foreground-cleaned map the dash-dot line, and the TOH
Wiener-filtered map the thin solid line.} \label{pcmpw}
\end{apjemufigure}

\begin{apjemufigure} 
\hbox{\hspace*{-0.5cm}
\centerline{\includegraphics[width=0.52\linewidth]{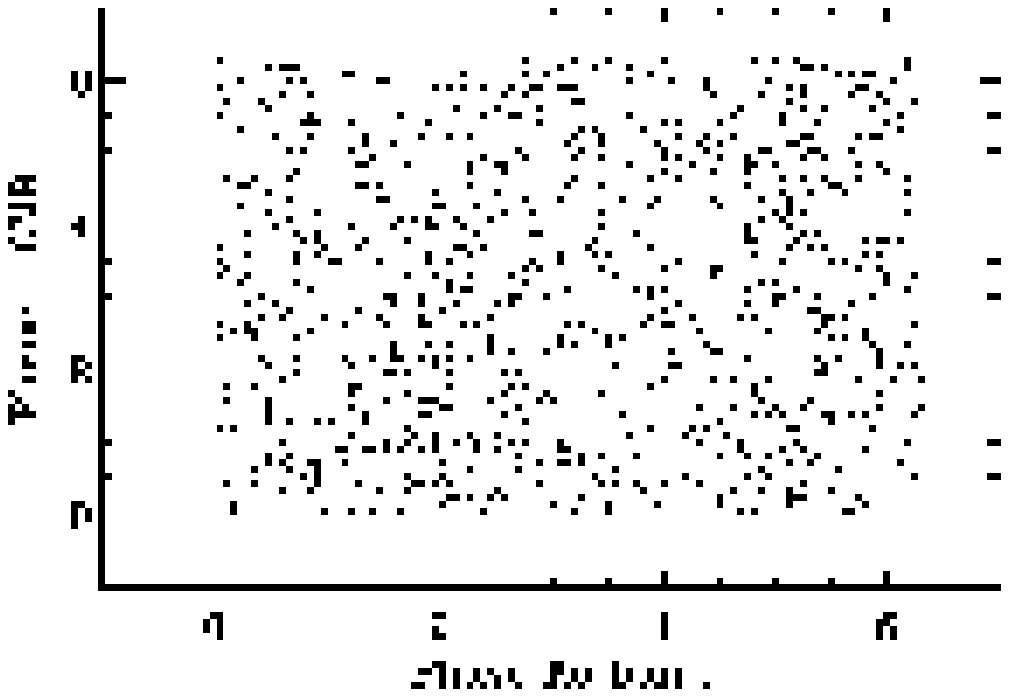}
 \includegraphics[width=0.52\linewidth]{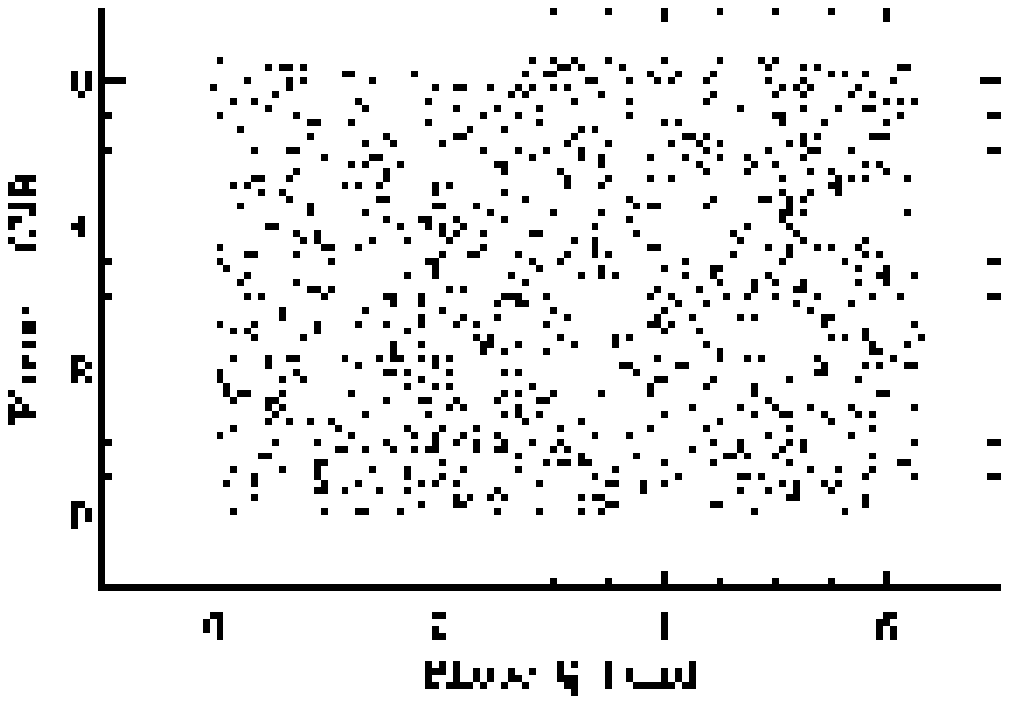}}}
\hbox{\hspace*{-0.5cm}
\centerline{\includegraphics[width=0.52\linewidth]{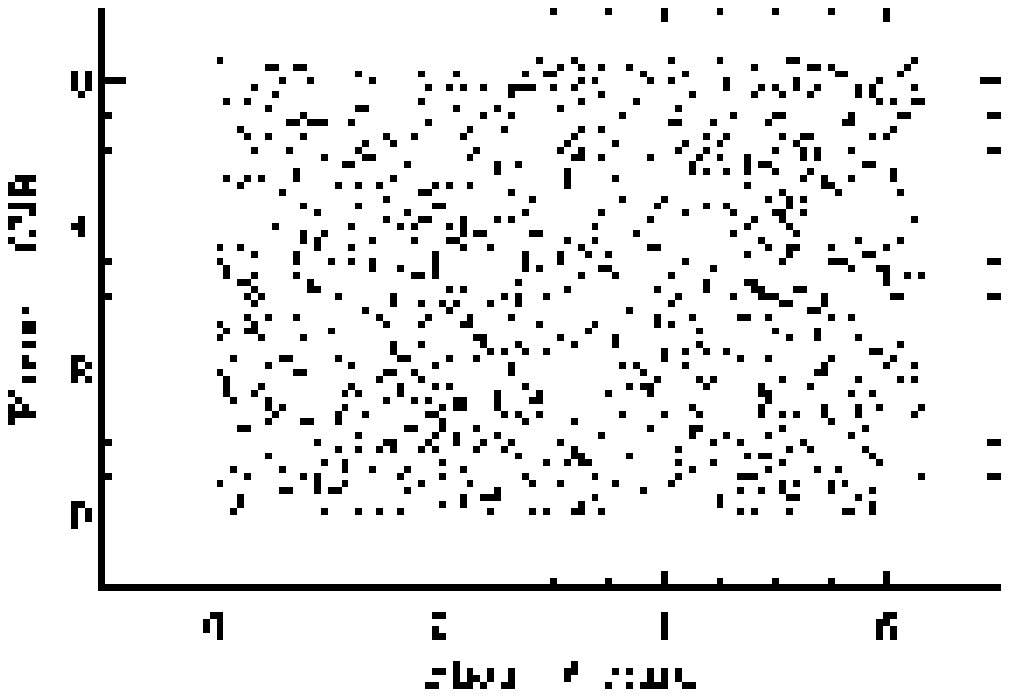}
\includegraphics[width=0.52\linewidth]{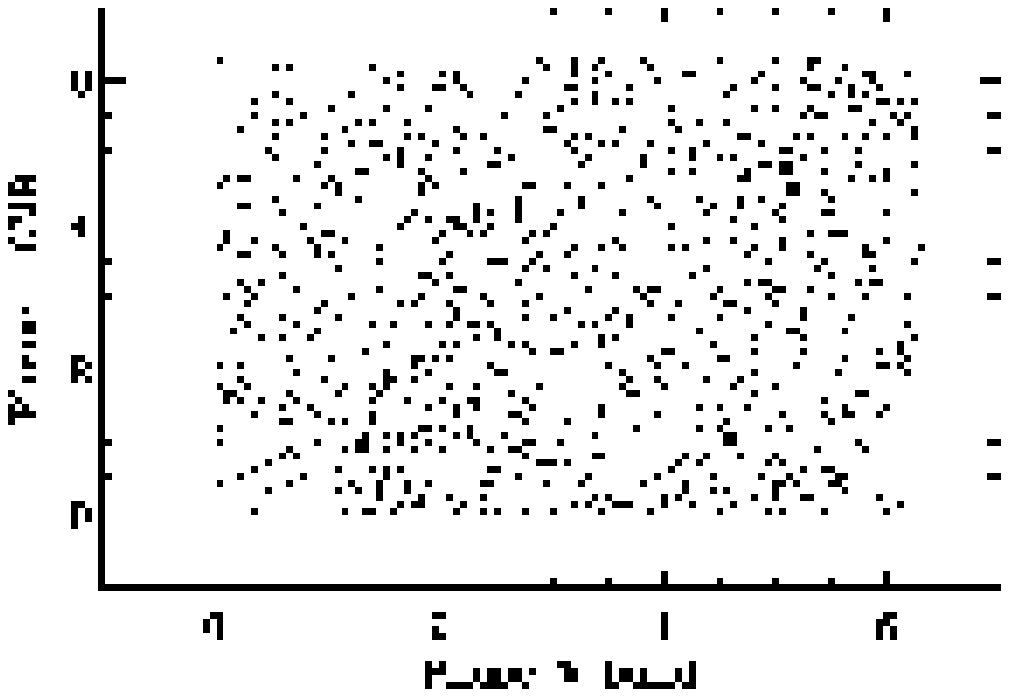}}}
\caption{Cross correlation of the phases between the PCM and the \wmap
foreground maps of the \wmap Ka, Q, V and W bands. Again the
scattering of the points shows that the phases of the PCM
practically have no correlations with those of the foregrounds.}
\label{pcmwmapxcorr}
\end{apjemufigure}

\begin{apjemufigure} 
\hbox{\hspace*{-0.5cm}
\centerline{\includegraphics[width=0.52\linewidth]{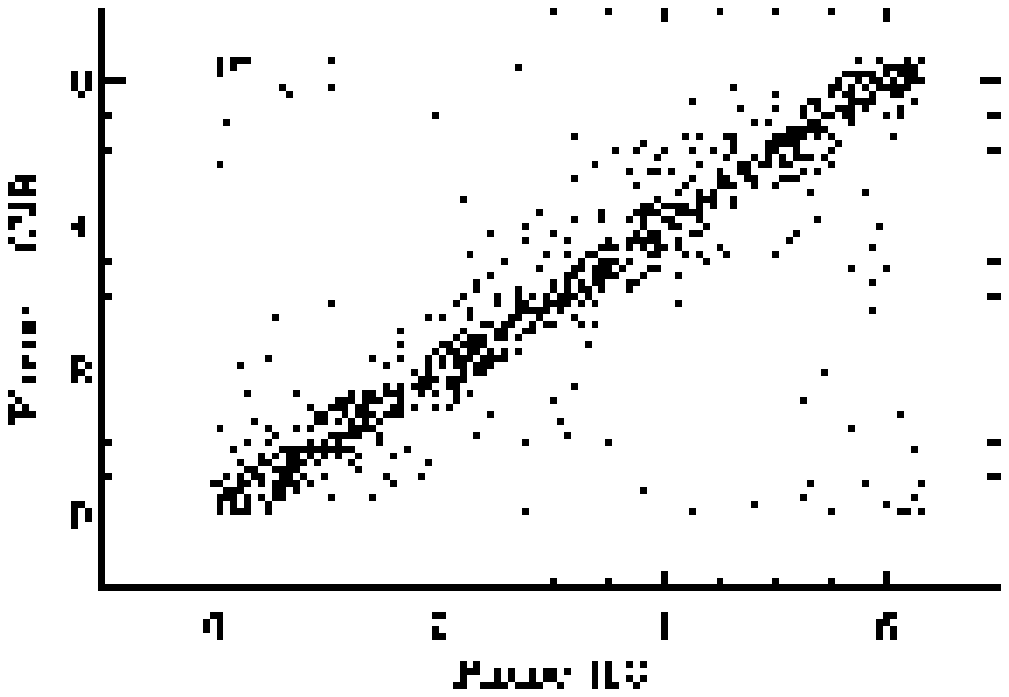}
 \includegraphics[width=0.52\linewidth]{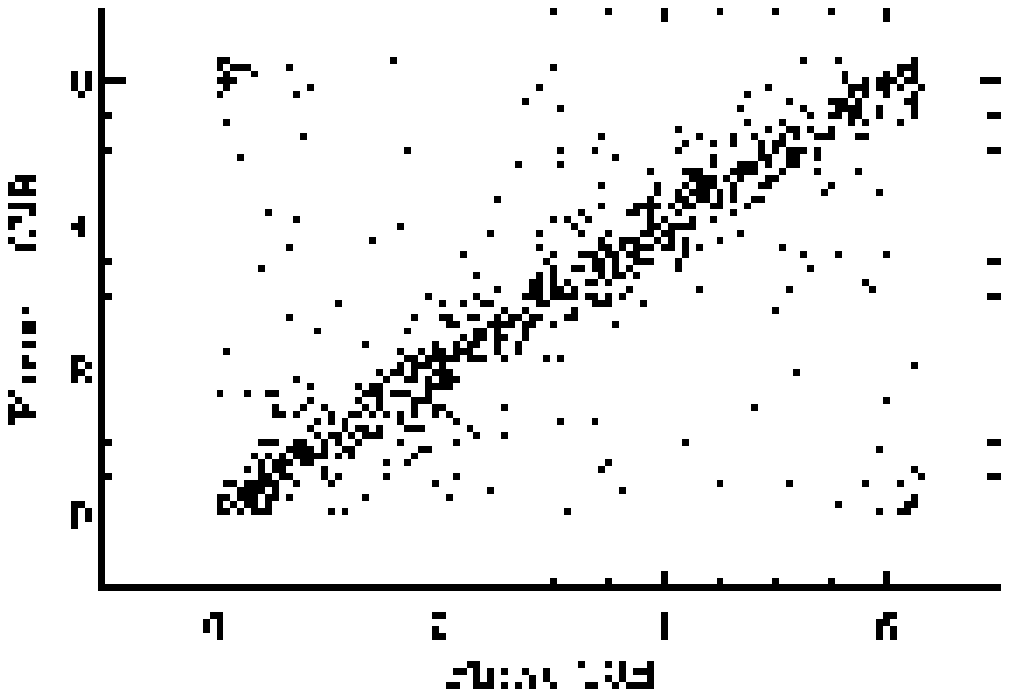}}}
\hbox{\hspace*{-0.5cm}
\centerline{\includegraphics[width=0.52\linewidth]{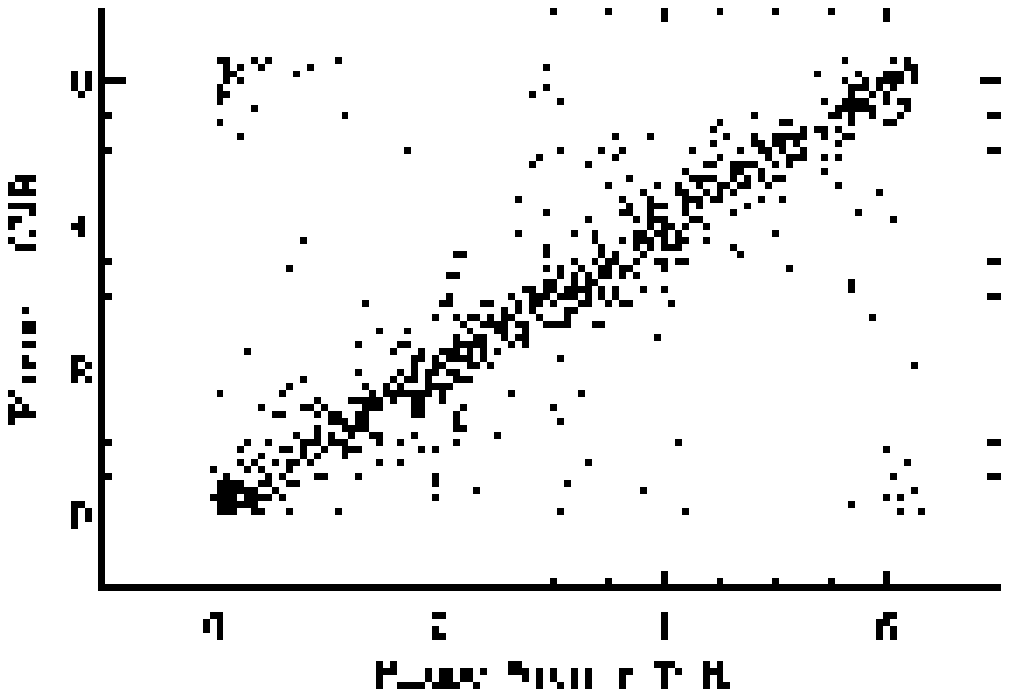}}}
\caption{Cross correlation of the phases between the PCM and the \wmap
ILC map (top left), between the PCM and the TOH foreground-cleaned map
(top right), and between the PCM and the TOH Wiener-filtered map (bottom).}
\label{pcmilcxcorr}
\end{apjemufigure}

Although the phases of the PCM have strong correlations with
those of the ILC map, the TOH foreground-cleaned map and the Wiener-filtered
map, as shown in Fig.\ref{pcmilcxcorr}, there are undoubtedly deviations
in phases between them, which is an indication of different morphology.  

The CMB map from the ``blind'' PCM method allows us
to reconstruct the foregrounds for each \wmap band.
In Fig.\ref{pcmforeground} we show at the top the foreground map at the 
V band, a simple subtraction of the PCM from the \wmap V-band map. In
the middle panel we show the \wmap
foreground map at the same band and the difference of these two panels is shown
at the bottom. As one can see from
this Figure, an additional power of the foregrounds from the PCM
subtraction than the \wmap foreground map is clearly shown in the
third panel, in particular we can see strong
contribution of the of the point-like sources residues along the
Galactic plane which are detected and removed
as foregrounds from the PCM subtraction.

\begin{apjemufigure} 
\hbox{\hspace*{-0.2cm}
\centerline{\includegraphics[width=1.0\linewidth]{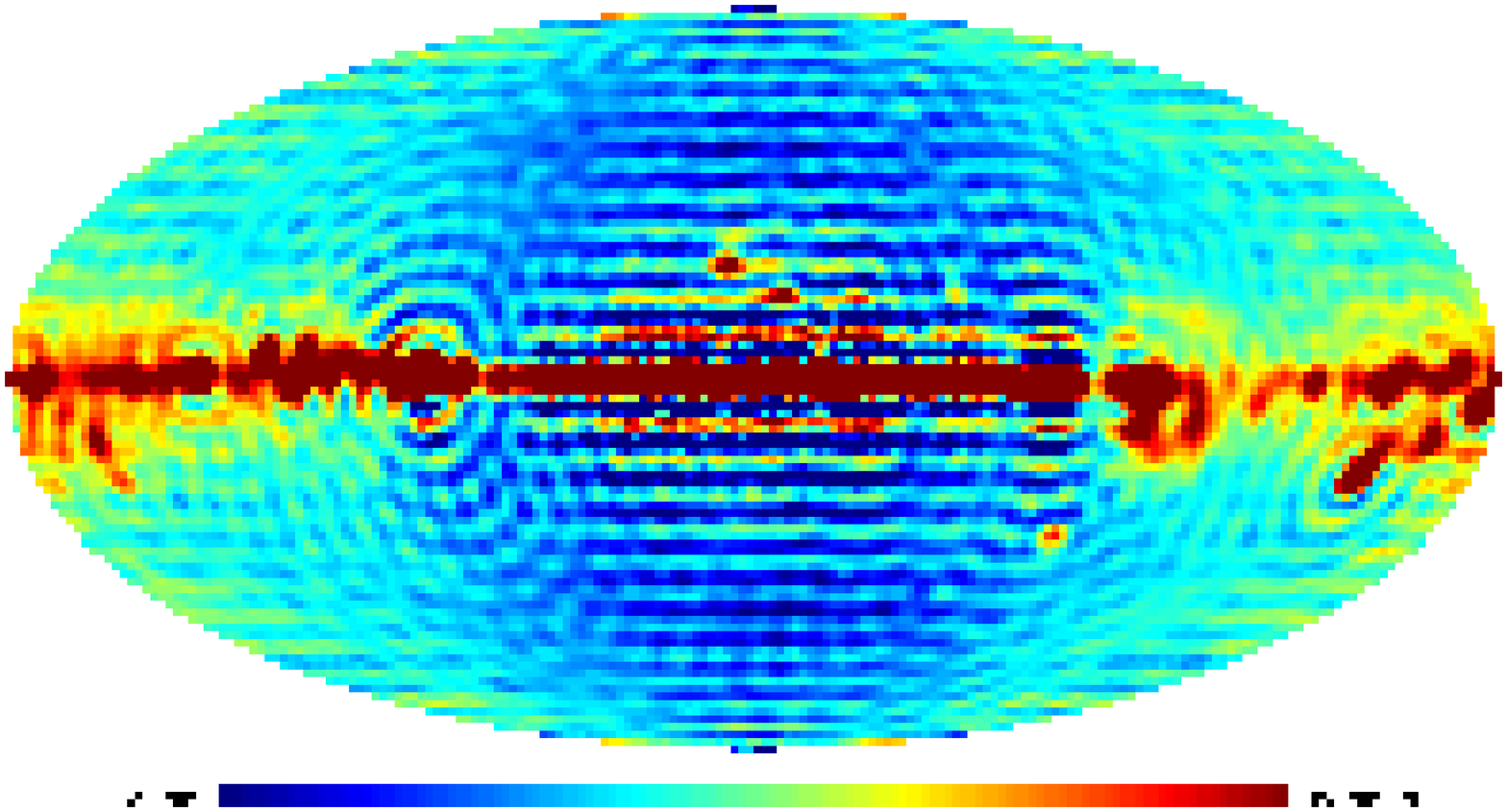}}}
\hbox{\hspace*{-0.2cm}
\centerline{\includegraphics[width=1.0\linewidth]{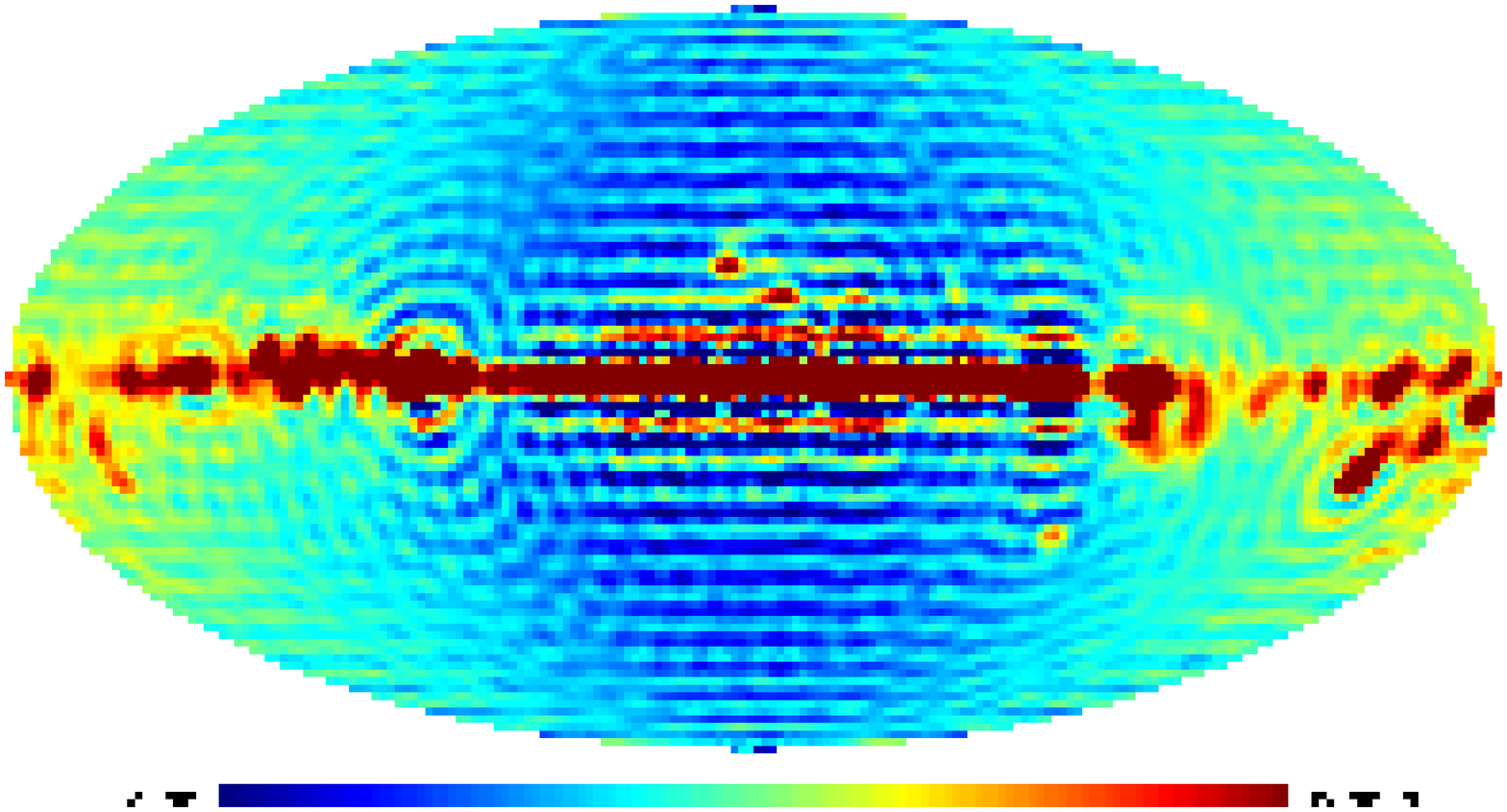}}}
\hbox{\hspace*{-0.2cm}
\centerline{\includegraphics[width=1.0\linewidth]{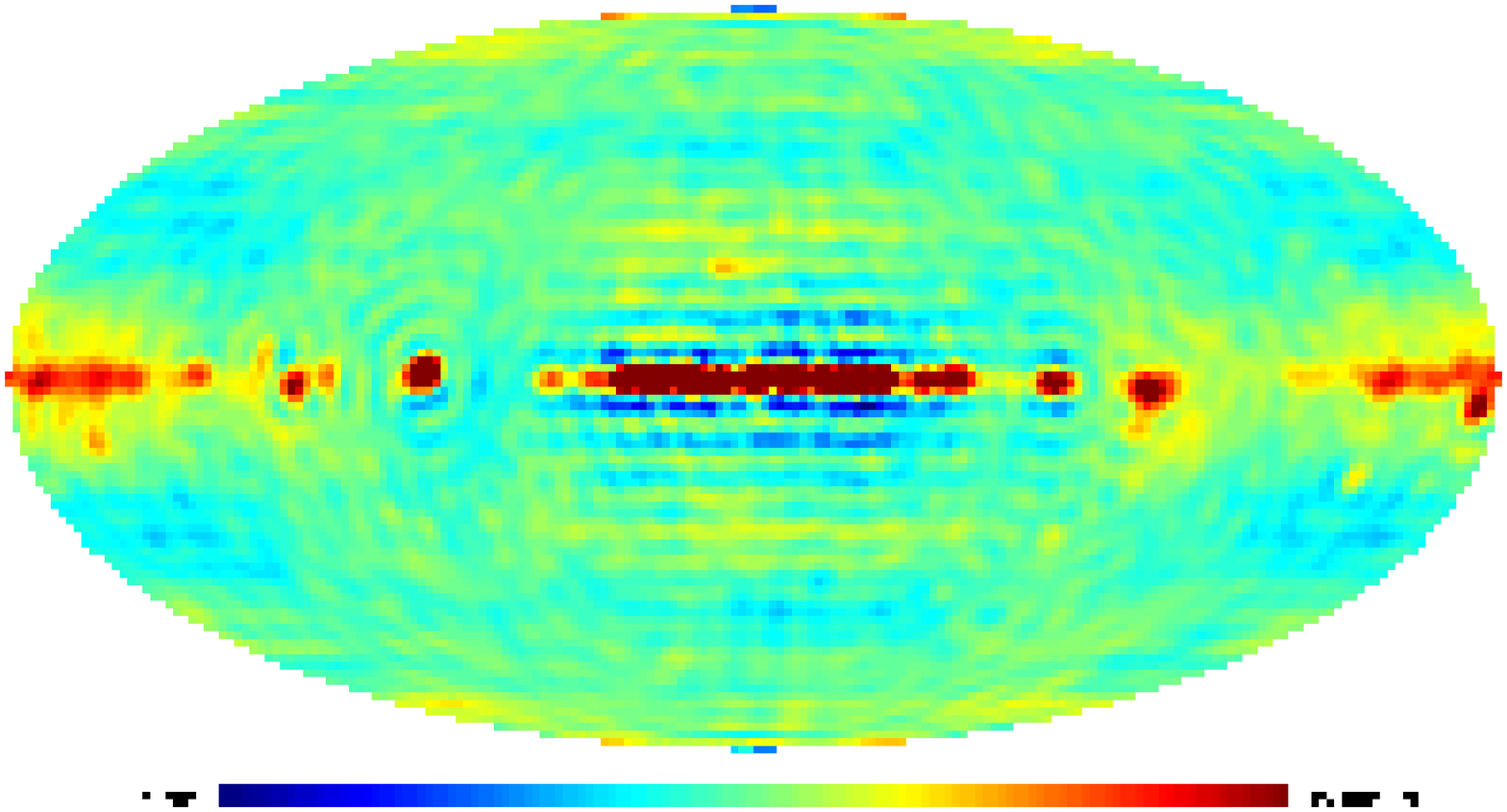}}}
\caption{The foreground map at the V band from the subtraction of the
PCM from the \wmap V-band map (top), the \wmap foreground map at the
V band (middle) and the difference of the first two maps (bottom).}
\label{pcmforeground}
\end{apjemufigure}

\section{Non-Gaussianity testing on the PCM}
In this section we use the phase
mapping technique as a non-Gaussian test to compare the non-Gaussianity
between the PCM, the \wmap ILC map and the TOH Wiener-filtered
map. This method has been successful in detecting non-Gaussianity in
the derived TOH maps from the \wmap data \citep{tacng}.

The null hypothesis of the phase mapping technique is based on the so-called
random-phase hypothesis, which is served as a practical definition of
homogeneous and isotropic Gaussian random fields \citep{bbks}. 
Therefore, any non-random phases from harmonic analyses is interpreted as
manifestation of non-Gaussianity. Due to the circular character of phases, we
use the return mapping of phases with certain separations $(\Delta m, \Delta
\ell)$ to render phases on return maps. Testing random phases is thus
equivalent to testing randomness of each return map. There are many ways
of quantifying such ensemble of return maps. We choose the simplest
$\meanchi$ statistic as in \citet{chisquare}: we first use
a Gaussian window function to smooth each return map and use a simple
$\meanchi =(1/M)\sum_{i,j} [ p(i,j)-\overline p ]^2/\overline p$
to yield the statistics, where $\overline p$ is the mean value and $M$
is the number of pixels. The null hypothesis will have mean
$\brameanchi_{\rm P} = 1 /4 \pi R^2$
and dispersion
$\Sigma_{\rm P}^2=1/\pi^3 R^2 (M/2)$.

We perform the mean $\chi^2$ statistics derived from phase mapping
technique for the PCM, the ILC map and the TOH Wiener-filtered map. We
take the phases from $\ell=2$ to maximum $\ell=40$ and render 256
points on each return map. The smoothing scale of the Gaussian window
function is $R=1$. The ensemble of return maps consists of separation
$\Delta m$ from $-9$ to 20 and $\Delta \ell$ from 1 to 15. From the
distribution function of such ensemble, we use the mean $\brameanchi$
and the dispersion $\Sigma$ of the distribution function as two
independent variables for displaying the statistics. We simulate 2000
realizations of Gaussian random fields to yields reliability.
In Fig.\,\ref{ng} the dashed and dotted contours indicate 68\%, 95\%
confidence level for detection, respectively. The non-Gaussianity through
phase mapping test shows that the PCM and the Wiener-filtered map have
close non-Gaussianity, whereas the \wmap ILC map shows less
non-Gaussian. Note that, however, both the ILC map
and Wiener-filtered map are produced by dissecting the whole-sky CMB
maps into 12 and 9 disjoint regions for cleaning, which systematically
change the phase configuration for the whole sky. 

\begin{apjemufigure} 
\hbox{\hspace*{-0.2cm}
\centerline{\includegraphics[width=1.0\linewidth]{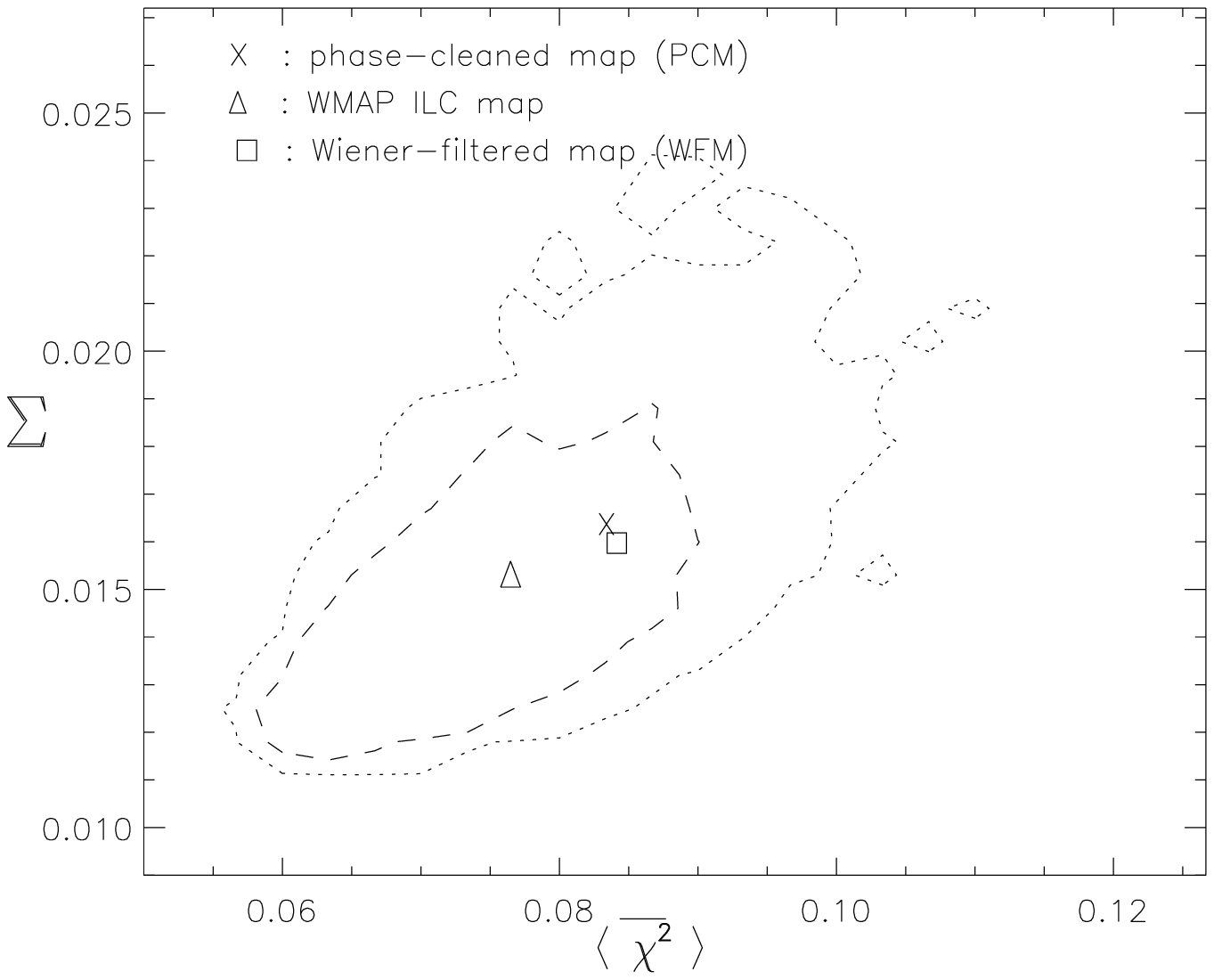}}}
\caption{Non-Gaussianity test using the phase mapping technique for the
PCM map (cross) at multipole range $2 \le \ell \le 40$, the \wmap ILC
map (triangle) and the TOH Wiener-filtered map (box). The dashed
contour corresponds to 68\% CL, the dotted contour corresponds to
95\% CL.}
\label{ng}
\end{apjemufigure}

\section{Discussions and conclusions}
In this paper we have proposed a new method for the separation of the
CMB signal from the Galactic and other foregrounds based on the phase
cleaning of the \wmap K--W bands at the multipole range $\ell \le
50$. This method is based on a simple but prevailing assumption that
the phases of the CMB signal should not correlate with those of the
foregrounds.  We first propose a ``non-blind'' PCM method using the
foreground maps available at the \wmap website as preliminary
foregrounds. We then introduce a ``blind'' PCM method with 4-step
foreground cleaning. We firstly group into pairs from the maps which
have strong cross correlations in phase. We then in step ${\bf b}$ use
a simple TOH minimization of the variance of the combination between
different pairs of K--W maps.
In the step ${\bf c}$ we propose a phase-cleaning filter for each
derived map from the constraint of minimal variance of phase
difference. Finally, we propose a MIN-MAX filter in order to obtain
the PCM map. Our CMB signal (the PCM) has the power spectrum in an excellent
agreement with the best fit \wmap cosmological model, but
the phases are slightly different from those of
the \wmap ILC map.
We also confirm that, after phase cleaning, the PCM has essentially
no correlation in phases with their cross-correlation with the
foregrounds. We would like to point out that the PCM method
can be used with different component separation methods such as the
Maximum Entropy Method (MEM) or another methods, which can be used
an initial step before the phase cleaning. Such generalization on the
PCM method is in progress.

Using the PCM method we have found corresponding corrections of the foreground
maps for each K--W band, which is useful for the data analysis
for the upcoming \planck mission.
The power spectra of the common foreground maps from our PCM
subtraction are slightly different from the \wmap foreground maps. Moreover,
the PCM does not show any serious contamination from the point-like
source residues mainly thanks to the MIN-MAX filter. 

The PCM CMB signal, as it follows from comparison between our CMB and
{\it WMAP} maps has a power smaller then ILC map.

We also confirm that the PCM displays a low level of power at the
quadrupole component, which also manifest itself in the \wmap ILC map
and the TOH map \citep{toh,oliveira}. However, the
low level in $\Cl$ is typical at the multipole range $2 \le \ell \le 10$.
In order to compare the power spectrum of the PCM with that of the \wmap best
fit model and the TOH Wiener-filtered map, in Fig.\ref{errorbars} we plot
these power spectra with the error bars caused by the cosmic variance.

\begin{apjemufigure} 
\hbox{\hspace*{-0.1cm}
\centerline{\includegraphics[width=0.9\linewidth]{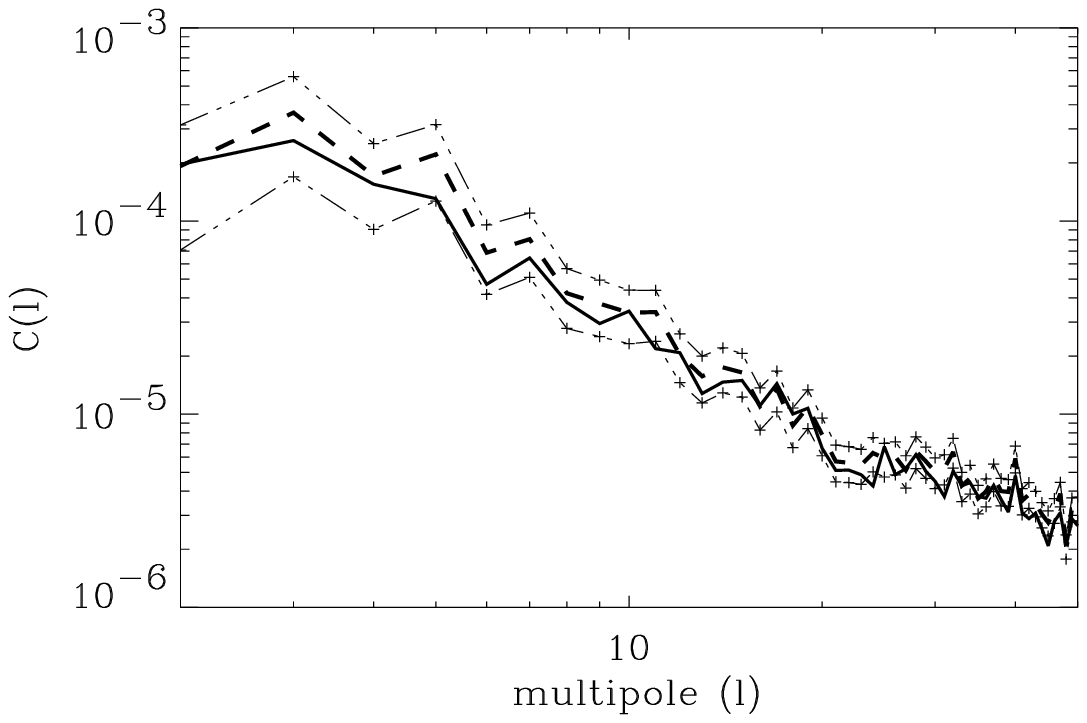}}}
\caption{The power spectra of the PCM (the thick solid line) and the
TOH Wiener-filtered map  (thick dashed line). The cosmic 
variance error bars with 68\% CL are indicated with crosses with the
triple dots dashed lines.}
\label{errorbars}
\end{apjemufigure}
As one can see from Fig.\ref{errorbars} all deviations between the PCM
and the TOH Wiener-filtered map are within the 68\% CL.
Some of the multipole range, however, seems to be quite
peculiar. Namely, at $20 \le \ell \le 24$
the power of both the PCM and the TOH Wiener-filtered map has some
suppression. Does it indicate that we have a peculiar realization
of a Gaussian random CMB field? Or we still have the contaminations of
the foregrounds which mimic as extra CMB power? All these
issues can only be discussed when more data are available, particularly
the future polarization observational data.

In this paper we concentrate only on the low multipole range of
the CMB $\ell \le 50$, at which the beam shape and the
instrumental noise are not crucial. The development of the PCM method
and the results at the multipole range $\ell > 50$ is in preparation.

\acknowledgements
This paper was supported by Danmarks Grundforskningsfond
through its support for the establishment of the Theoretical
Astrophysics Center. We thank M. Tegmark et al. for providing
their processed maps. We thank G. Efstathiou, A. Lasenby, A. Doroshkevich,
M. Demianski and P. R. Christensen for useful discussions.
We acknowledge the use of \healpix
\footnote{\tt http://www.eso.org/science/healpix/}
package \citep{healpix} to produce $\alm$ from the \wmap data and the
use of the \glesp package \citep{glesp} for data analyses and the
whole-sky figures.


\appendix
\section{A : The concept of the ``blind'' PCM method}
From Eq.(\ref{eq9}) and (\ref{eq12}), one might think that it is
possible to generalize the ``non-blind'' phase reconstruction method
to the ``blind'' variant simply using the $\alm$ coefficients in
Eq.(\ref{eq12}) as an estimator for the reconstructed phases of the
pre-CMB signal. Such generalization, however, would be incorrect
because of the following reason. Below we consider only for the
multipole range $\ell \le 50$ to avoid the instrumental noise and beam
shape properties. According to Eq.(\ref{eq5}) the moduli of the $\alm$
coefficients are related with the moduli of the foregrounds and CMB
for each \wmap band  as (with the phases $\Ks^{(j)}$ in Eq.(\ref{eq5})
being replaced by $\Ph^{(j),f}$, the resultant foreground phases) 
\begin{equation}
|\a^{(j)}|^2=|\G^{(j)}|^2 +
2|\G^{(j)}||\a^{cmb}|\cos(\Ph^{(j),f}-\ks)+|\a^{cmb}|^2  \label{a1}
\end{equation}
where $|\a^{cmb}| \ll |\G^{(j)}|$. Using Taylor series one can obtain
$|\a^{(j)}| \sim |\G^{(j)}|+|\a^{cmb}|\cos(\Ph^{(j),f}-\ks)$. Then, from
Eq.(\ref{eq11}) we get 
\begin{eqnarray}
\sum_j(-1)^{j+1} \w^{(j)}|\a^{(j)}| & \sim &
 \frac{|\a^{cmb}|}{|\a^{(1)}-\a^{(2)}|}
 \left[|\G^{(1)}|\cos(\Ph^{(1),f}-\ks)-
 |\G^{(2)}|\cos(\Ph^{(2),f}-\ks)\right]  \nonumber \\
&=& \frac{|\a^{cmb}|}{|\a^{(1)}-\a^{(2)}|}\cos(\Ph^{(1),f}-\ks)\left[
|\G^{(1)}|-|\G^{(2)}|\cos\Delta+|\G^{(2)}|\sin\Delta \, \tan(\Ph^{(1),f}-\ks)
  \right] 
\label{a2} 
\end{eqnarray}
The comparison of Eq.(\ref{a2}) and Eq.(\ref{eq11}) allow us to conclude
that the deviation of the reconstructed phases from the CMB phases should be
very high  even for the small values of $\Delta$-parameter. This means 
that the reconstructed phases of the CMB signal would be close to the
foreground phases and the corresponding power spectrum should have very
significant errors.
It is important to note that such a conclusion means that in order
to extract the correct phases and amplitudes of the CMB signal from the
data using a ``blind'' method of separation, it is necessary to
decrease the contamination in Eq.(\ref{a1}) of the linear term
($\sim|\a^{cmb}|\cos(\Ph^{(j),f}-\ks)$). The only way to do this is to
take into account that $\sum_m|\a^{cmb}|\cos(\Ph^{(j),f}-\ks)
\rightarrow 0$  for
high $\ell,m$. In such a case $\sum_j(-1)^{j+1} \w^{(j)}|\a^{(j)}|\sim
{\cal O}(|\a^{cmb}|^2)$ and corresponding error of the CMB phase
reconstruction should be of the order
$|\a^{cmb}|/\min\{|\G^{(j)}| \}$. Moreover, averaging over the $m$ values
means that the corresponding weighting coefficients $\w^{(j)}$ is a
function of $\ell$ only, but not of $m$, which is ideologically close to
\citet{te96} and the TOH method, where
$\w^{(j)}=w^{(j)}(\ell)$. We would like to point
out, however, that for such an optimization, as indicated in
Eq.(\ref{eq11}), the reconstructed CMB phases will have the
cross correlations with the foregrounds phases at different levels for
different modes.  \\
\vspace{1cm}

\section{B : Minimization of the phase difference}
Assuming that the reconstructed phase $\Ph^{M}$ should be close to the
CMB phase $\ks$, from Eq.(\ref{eq11}), we
obtain 
\begin{equation}
\Ph^{M}= \ks + \arcsin\frac{\sum_j
\g^{(j)}|\G^{(j)}|\sin({\Ph^{(j),f} -\ks)}\cos \Ph^{M}} {\sum_j
\g^{(j)}|\G^{(j)}|\cos{\Ph^{(j),f}} + |\a^{cmb}|\cos{\ks}}. \label{b1}
\end{equation}
As mentioned above, a new element of the ``blind'' reconstruction
of the CMB signal is the correlation between the
derived phases $\Ph^{M}$ and the foreground phases $
\Ph^{(j),f}_1,\Ph^{(j),f}_2$ for all linear combinations of the
maps. Moreover, if 
\begin{equation}
 |\a^{cmb}|\cos{\ks}\gg\sum_j  \g^{(j)}|\G^{(j)}|\cos{\Ph^{(j),f}}, \label{b2}
\end{equation}
then 
$\Ph^{M} \simeq \ks$ while for some of the $\ell,m$ modes which corresponds to inequality
$|\a^{cmb}| \cos{\ks} \ll \sum_j \g^{(j)}|\G^{(j)}|\cos{\Ph^{(j)}}$, 
reconstruction of the CMB phases needs more specific investigation,
which will be discussed at the end of the section.
For all $\ell,m$ modes satisfying Eq.(\ref{b2})
in order to decrease the correlation between the foregrounds and
CMB phases we introduce
the optimization scheme based on the iterative solution  of the Eq.(\ref{b1}).
The first step of iterations is to minimize the variance 
\begin{equation}
V=\frac{1}{2\pi}\int_{0}^{2\pi}d\ks\left(\Ph^M-\ks\right)^2
\rightarrow \min, \label{b3}
\end{equation}
which corresponds to the equation 
\begin{equation}
\frac{\delta V}{\delta \g^{(k)}}=\frac{1}{4{\pi}}\frac{\delta}{\delta
\g^{(k)}} \left\{\sum_{i,j}  \frac{\g^{(i)} \g^{(j)}
  |\G^{(j)}||\G^{(i)}|}{ |\a^{cmb}|^2}
\left[I_1\cos(\Ph^{(j),f} -\Ph^{(i),f})-I_2\right] \right\}=0  ,
		\label{b4}
\end{equation}
where $\delta/\delta \g^{(k)}$ are the functional derivatives and
\begin{eqnarray}
 I_1& = &\int_0^{2\pi}d\ks\frac{\cos^2\ks}{(\cos\ks +\beta)^2}, \nonumber \\
 I_2& = & \int_0^{2\pi}d\ks\frac{\cos^2\ks \cos(\Ph^{(j),f}+\Ph^{(i),f} -
 2\ks)}{(\cos\ks +\beta)^2} , \label{b5}
\end{eqnarray}
and 
\begin{equation}
\beta=\frac{1}{|\a^{cmb}|} \sum_j  \g^{(j)}|\G^{(j)}|\cos{\Ph^{(j),f}}.
\label{b6}
\end{equation}
Note that $\beta$ could be both positive and negative depending on
the phases ${\Ph^{(j),f}}$.
As we mention above,  Eq.(\ref{b4}) has an exact solution not
for all values of $\ell,m$.
For $\ell,m$ satisfying Eq.(\ref{b2}) we can neglect the $\beta$ parameter
in Eq.(\ref{b5}) and obtain $I_1 \simeq 2\pi$, $I_2 \simeq 0$.
Thus, from Eq.(\ref{b4})-(\ref{b6}) we have
\begin{eqnarray}
\g^{(1)} & = & \frac{|\G^{(2)}|\left[|\G^{(2)}|-
|\G^{(1)}|\cos(\Ph^{(2),f}-\Ph^{(1),f})\right]}{|\G^{(1)}-\G^{(2)}|^2},
\nonumber \\
\g^{(2)} & = & \frac{|\G^{(1)}|\left[|\G^{(1)}|-|\G^{(2)}|
\cos(\Ph^{(2),f}-\Ph^{(1),f})\right]}{|\G^{(1)}-\G^{(2)}|^2}. \label{b7}
\end{eqnarray}
The filter Eq.(\ref{b7}) depends on both $\ell$ and $m$. As one can
see from Eq.(\ref{b7}), if the phases of the foregrounds $\Ph^{(2),f}$
and $\Ph^{(1),f}$ are identical, then
one can have the same form for the  $\g^{(1)}$ and $\g^{(2)}$ weighting
coefficients as in Eq.(\ref{eq12}).
Unfortunately, because of the non-zero phase difference
$\Ph^{(2),f}-\Ph^{(1),f}$ 
the filter (\ref{b7}) is not useful for the ``blind'' method of the
CMB reconstruction. It is due to the instability of the
iteration sheme, which reproduces at the very first step the phases of the
foregrounds and use them as an attractor.
 
Now following the concept in Appendix A, let us introduce another
concept of the phase optimization based on the minimization of the
weighting phase variance
\begin{equation}
V=\frac{1}{2 \pi (2\ell+1)}\sum_m\frac{|\a^{cmb}|^2}{\Cl}\int_{0}^{2\pi}d\ks\left(\Ph^M-\ks\right)^2\rightarrow \min,		\label{b8}
\end{equation}
which corresponds to the following
\begin{equation}
\frac{\delta V}{\delta \g^{(k)}} \propto \frac{\delta}{\delta \g^{(k)}}\left[ 
\sum_m \sum_{i,j}  \g^{(i)} \g^{(j)} |\G^{(j)}||\G^{(i)}|
\cos(\Ph^{(j),f} -\Ph^{(i),f}) \right]=0, \label{b9}
\end{equation}
and  
\begin{eqnarray}
\g^{(1)} & = & \frac{\sum_m |\G^{(2)}|\left[\,|\G^{(2)}|-
|\G^{(1)}|\cos(\Ph^{(2),f}-\Ph^{(1),f})\,\right]}{\sum_m
|\G^{(1)}-\G^{(2)}|^2}, \nonumber \\
\g^{(2)} & = & \frac{\sum_m |\G^{(1)}|\left[\,|\G^{(1)}|-|\G^{(2)}|
\cos(\Ph^{(2),f}-\Ph^{(1),f})\,\right]}{\sum_m
|\G^{(1)}-\G^{(2)}|^2}. \label{b10} 
\end{eqnarray}

It is easy to demonstrate that the filter Eq.(\ref{b10}) corresponds
to the minimum of the variance $\sum_m |\a^M-\a^{cmb}|^2$. We propose
to use this filter for optimization of the pre-CMB reconstruction at
the second and subsequent iterations. \\
\vspace{1cm}

\section{C : The ``$\pi/2$'' problem of the phases reconstruction}
In this Appendix we discuss some problems of the CMB image reconstruction
related with optimization methods. The exact solution for the reconstructed
phases in a form of Eq.(\ref{b10}) has
some peculiar reconstructed phases $\Ph^M$ if the CMB phases $\ks$ are close to  $\pi/2$ and $3\pi/2$.
In such a case, from Eq.(\ref{b4}) we obtain
\begin{equation}
\Ph^M \simeq \tan^{-1}\left(\frac{|a^{cmb}| \cos\delta}{\sum_j
  \g^{(j)}|\G^{(j)}|\cos{\Ph^{(j),f}}}+
  \frac{\sum_j\g^{(j)}|\G^{(j)}|\sin{\Ph^{(j),f}}}{\sum_j
  \g^{(j)}|\G^{(j)}|\cos{\Ph^{(j),f}}}\right),  
\end{equation}
where $\ks=\pi/2+\delta$,
$\delta \ll 1$ and $|a^{cmb}| \cos\ks\ll\sum_j
 \g^{(j)}|\G^{(j)}|\cos{\Ph^{(j),f}}$.
We point out that depending on the foreground phases the
reconstructed phase $\Ph^M$ could be arbitrary.
It means that phases close to $\pi/2$ will be reconstructed with a big
error. The same argument can be applied to the case when $\ks$ is close
to $3\pi/2$.

\end{document}